\documentclass[12pt]{article}

\usepackage{amsmath,amssymb,amsfonts,amsthm,hyperref,bbm,latexsym,epsfig}
\usepackage{graphicx,multirow}
\usepackage{caption}

\usepackage[dvips]{pstcol}
\definecolor{LemonChiffon}{rgb}{1.,0.98,0.8}
\definecolor{LightGray}{gray}{0.85}
\definecolor{Blue}{rgb}{0.,0.,1.}
\definecolor{Pink}{rgb}{1.,0.75,0.8}
\definecolor{Gold}    {rgb}{1.,0.84,0.}
\definecolor{RedA}{hsb}{0.9,0.3,0.7}
\definecolor{RedB}{hsb}{0.9,0.3,1}
\definecolor{Beige}   {rgb}{0.96,0.96,0.86}
\definecolor{MyYellow}{rgb}{1.,0.84,0.8}
\definecolor{ColorA}   {hsb}{0.2,0.2,1}
\definecolor{ColorB}   {hsb}{0.2,0.2,0.8}
\definecolor{LightCyan}{rgb}{0.88,1.,1.}

\newcommand{\bs}{\begin{subequations}}
\newcommand{\es}{\end{subequations}}
\newcommand{\be}{\begin{equation}}
\newcommand{\ee}{\end{equation}}
\newcommand{\ba}{\begin{eqnarray}}
\newcommand{\ea}{\end{eqnarray}}
\newcommand{\no}{\nonumber \\}

\allowdisplaybreaks

\begin{document}

\title{
\normalsize \hfill CFTP/18-009
\\[6mm]
\LARGE The three- and four-Higgs couplings in the \\
general two-Higgs-doublet model}

\author{
\addtocounter{footnote}{2}
D.~Jur\v{c}iukonis$^{(1)}$\thanks{\tt darius.jurciukonis@tfai.vu.lt}
\ {\normalsize and}
L.~Lavoura$^{(2)}$\thanks{\tt balio@cftp.tecnico.ulisboa.pt}
\\*[3mm]
$^{(1)} \! $
\small University of Vilnius,
\small Institute of Theoretical Physics and Astronomy, \\
\small Saul\.{e}tekio ave.~3, LT-10222 Vilnius, Lithuania 
\\[2mm]
$^{(2)} \! $
\small Universidade de Lisboa, Instituto Superior T\'ecnico, CFTP, \\
\small 1049-001 Lisboa, Portugal
\\*[2mm]
}

\date{\today}

\maketitle

\begin{abstract}
  We apply the unitarity bounds and the bounded-from-below (BFB) bounds
  to the most general scalar potential of the two-Higgs-doublet model (2HDM).
  We do this in the Higgs basis,
  \textit{i.e.}\ in the basis for the scalar doublets
  where only one doublet has vacuum expectation value.
  In this way we obtain bounds on the scalar masses
  and couplings that are valid for all 2HDMs.
  We compare those bounds to the analogous bounds
  that we have obtained for other simple extensions of the Standard Model (SM),
  namely the 2HDM extended by one scalar singlet
  and the extension of the SM through two scalar singlets.
\end{abstract}

\newpage

\section{Introduction}

In order to unveil the detailed mechanism of electroweak symmetry breaking
it is crucial to measure the self-couplings of the boson with mass 125\,GeV
discovered in 2012 at the LHC~\cite{discovery}.
In this paper we call that boson $h_1$.
The Standard Model (SM) predicts $h_1$ to be a scalar
and predicts its cubic and quartic couplings $g_3$ and $g_4$,
which we define through
\be
\label{kmhiho}
\mathcal{L} = \cdots - g_3 \left( h_1 \right)^3 - g_4 \left( h_1 \right)^4,
\ee
to be $g_3^\mathrm{SM} \approx 32$\,GeV and $g_4^\mathrm{SM} \approx 0.032$,
respectively.
However,
in Nature the scalar sector may be more complicated
than in the SM~\cite{Ivanov:2017dad}
and then $g_3$ and $g_4$ might have very different values.
In this paper we survey the allowed values of $g_3$ and $g_4$
in three extensions of the SM:
\begin{itemize}
\item The SM plus two real,
  neutral scalar singlets and with a reflection symmetry
  on each of those singlets.
  Let SM2S denote this model,
  which we treat in section~\ref{sec:singlets}.
\item The two-Higgs-doublet model (2HDM),
  which is the focus object of section~\ref{sec:2HDM}.
\item The 2HDM with the addition of one real,
  neutral scalar singlet
  and with a reflection symmetry of that singlet.
  This model,
  which we dubb the 2HDM1S,
  is dealt with in section~\ref{sec:2HDMsinglet}.
\end{itemize}
Our ingredients for bounding $g_3$ and $g_4$ in each of these models are:
\begin{itemize}
\item The bounded-from-below (BFB) and the unitarity conditions
  on the quartic part of the scalar potential of each model.
  We apply those conditions directly in the basis for the scalar doublets
  where only one of them has vacuum expectation value (VEV).
\item The experimental bound on the oblique parameter $T$~\cite{RPP}.
\item The (approximate) bound $\cos{\vartheta} > 0.9$
  on the $h_1$ component $\cos{\vartheta}$
  of the scalar doublet with nonzero VEV.
\end{itemize}
Other authors before us~\cite{baglio}--\cite{DM1}
have used the BFB and unitarity constraints
in order to bound the scalar masses and couplings of the 2HDM.
However,
they have done it in the context of a constrained version of the model,
\textit{viz.}\ the 2HDM with a reflection symmetry
acting on one of the scalar doublets,
leading to $\lambda_6 = \lambda_7 = 0$ in the scalar potential
of equation~\eqref{gjihoree}.
In this paper we deal on the fully general 2HDM.
We enforce the BFB and unitarity constraints in the so-called Higgs basis,
\textit{i.e.}\ the basis where only one of the doublets has VEV.
Since that basis exists for every 2HDM,
we thus obtain results that apply to every 2HDM.
  
At present there are only indirect,
very rough bounds on $g_3$.
Using the Standard Model Effective Theory developed in ref.~\cite{haisch}
and experimental data~\cite{ATLAS},
ref.~\cite{rindani} has found that
$-8.4 < g_3 \left/ g_3^\mathrm{SM} \right. < 13.4$. From
the contribution of $g_3$ to the oblique parameters $S$ and $T$,
ref.~\cite{kribs} derived $-14.0 < g_3 \left/ g_3^\mathrm{SM} \right. < 17.4$.
The authors of ref.~\cite{maltoni1} obtained firstly
$-9.4 < g_3 \left/ g_3^\mathrm{SM} \right. < 17.0$
and then~\cite{maltoni2}
$-8.2 < g_3 \left/ g_3^\mathrm{SM} \right. < 13.7$.
The partial-wave unitarity of $h_1 h_1 \to h_1 h_1$ scattering
has been used~\cite{luzio} to obtain
$\left| g_3 \left/ g_3^\mathrm{SM} \right. \right| \lesssim 6.5$
and $\left| g_4 \left/ g_4^\mathrm{SM} \right. \right| \lesssim 65$.
In an analysis of a specific three-Higgs-doublet model,
ref.~\cite{3HDM} has found that in that model
$-1.3 < g_3 \left/ g_3^\mathrm{SM} \right. < 20.0$
and $1.05 < g_4 \left/ g_4^\mathrm{SM} \right. < 1.6$.

The measurement of $g_3$ should be possible
at future colliders, and may even
  eventually become possible at the LHC~\cite{LHC}.
Reference~\cite{divita} concluded that
one may be able to measure $g_3$ provided
$-0.72 < g_3 \left/ g_3^\mathrm{SM} \right. < 7.05$.
Unfortunately,
measuring $g_4$ is probably more challenging~\cite{plehn}.

\subsection{$g_3$ and $g_4$ in the SM}

The Standard Model has only one scalar doublet $\phi_1$.
We write it
\be
\label{nuiho}
\phi_1 = \left( \begin{array}{c} G^+ \\ v + \left( H + i G^0 \right)
  \left/ \sqrt{2} \right. \end{array} \right),
\ee
where $v$ is the VEV,
which is real and positive,
and $G^+$ and $G^0$ are (unphysical) Goldstone bosons.
In the SM $H$ coincides with the observed scalar $h_1$.
The scalar potential is
\be
V = \mu_1 \phi_1^\dagger \phi_1
+ \frac{\lambda_1}{2} \left( \phi_1^\dagger \phi_1 \right)^2.
\ee
The minimization condition of $V$ is $\mu_1 = - \lambda_1 v^2$.
Therefore,
in the unitary gauge where $G^\pm$ and $G^0$ do not exist,
\be
\label{fuiho}
V = - \frac{\lambda_1 v^4}{2} + \lambda_1 v^2 H^2
+ \frac{\lambda_1 v}{\sqrt{2}}\, H^3 + \frac{\lambda_1}{8}\, H^4.
\ee
The second term in the right-hand side of equation~\eqref{fuiho}
indicates that the squared mass $M_1$ of the observed scalar
is given by $M_1 = 2 \lambda_1 v^2$.
Therefore,
\bs
\label{djikn3}
\ba
V &=& - \frac{M_1 v^2}{4} + \frac{M_1}{2} \left( h_1 \right)^2
+ \frac{M_1}{2 \sqrt{2} v} \left( h_1 \right)^3
+ \frac{M_1}{16 v^2} \left( h_1 \right)^4
\label{cuigho} \\ &=&
\cdots + g_3^\mathrm{SM} \left( h_1 \right)^3
+ g_4^\mathrm{SM} \left( h_1 \right)^4.
\label{djikn}
\ea
\es
Using the approximate experimental values
\bs
\label{data}
\ba
\label{M1}
M_1 &=& \left( 125\, \mathrm{GeV} \right)^2,
\\
\label{v}
v &=& 174\,\mathrm{GeV},
\ea
\es
one gathers from equation~\eqref{cuigho} that
\bs
\label{fnjkno}
\ba
g_3^\mathrm{SM} = \frac{M_1}{2 \sqrt{2} v} &=& 31.7\,\mathrm{GeV}, \label{g3}
\\
g_4^\mathrm{SM} = \frac{M_1}{16 v^2} &=& 0.0323.
\ea
\es
It should be noted that the sign of $g_3$
implicitly depends on the sign of $h_1$.
We fix that sign by noting that the covariant derivative of $\phi_1$
gives rise to a term
\bs
\ba
\mathcal{L} &=& \cdots + \frac{g^2}{2}\, W_\mu^+ W^{\mu -}
\left( v + \frac{H}{\sqrt{2}} \right)^2
\\ &=& \cdots +  \frac{g^2 v}{\sqrt{2}}\ W_\mu^+ W^{\mu -} H.
\label{cbuiho}
\ea
\es
Thus,
the coupling $W_\mu^+ W^{\mu -} h_1$,
\textit{viz.}\ $g^2 v \left/ \sqrt{2} \right.$,
is positive.

\section{The Standard Model plus two singlets}
\label{sec:singlets}

We consider the Standard Model with the addition of two real
$SU(2) \times U(1)$-invariant scalar fields
$S_1$ and $S_2$.\footnote{In appendix~\ref{appendixA}
    we treat the simpler case of the HSM,
    \textit{viz.}\ the Standard Model with the addition of
    \emph{only one}\/ real gauge singlet.}
We assume two symmetries $S_1 \to - S_1$ and $S_2 \to - S_2$.
We call this model the SM2S.\footnote{The
SM2S has already been mentioned in the literature as a model for Dark Matter,
see ref.~\cite{DM2}.}
The scalar potential is
\bs
\label{potpotpot}
\ba
V &=& V_2 + V_4, \label{hoihp} \\
V_2 &=& \mu_1 \phi_1^\dagger \phi_1 + m_1^2 S_1^2 + m_2^2 S_2^2, \\
V_4 &=& \frac{\lambda_1}{2} \left( \phi_1^\dagger \phi_1 \right)^2
+ \frac{\psi_1}{2}\, S_1^4
+ \frac{\psi_2}{2}\, S_2^4
+ \psi_3 S_1^2 S_2^2
+ \phi_1^\dagger \phi_1 \left( \xi_1 S_1^2 + \xi_2 S_2^2 \right).
\hspace*{6mm}
\ea
\es

\subsection{Unitarity condidions}

We derive the unitarity conditions on the parameters
of $V_4$.\footnote{Strictly speaking,
  the unitarity conditions derived and utilized in this paper
  are the ones valid in the limit of infinite Mandelstam parameter $s$.
  For finite $s$ one must take into account the trilinear vertices
  that are induced from the quartic vertices
  when one substitutes one of the fields by its VEV.
  The unitarity conditions then become $s$-dependent
  and may be either more or less restrictive than the conditions
  in the limit of infinite $s$.
  See ref.~\cite{staub}.}
We follow closely the method of ref.~\cite{silva}.
We write
\be
\phi_1 = \left( \begin{array}{c} a \\ b \end{array} \right),
\quad
\phi_1^\dagger = \left( \begin{array}{cc} a^\ast & b^\ast \end{array} \right),
\quad
S_1^\ast = S_1, \quad S_2^\ast = S_2,
\ee
where $a$ and $b$ are complex fields.
Then,
\bs
\ba
V_4 &=& \frac{\lambda_1}{2} \left( a^\ast a^\ast a a
+ b^\ast b^\ast b b + 2 a^\ast b^\ast a b \right)
+ \frac{\psi_1}{2}\, S_1^4
+ \frac{\psi_2}{2}\, S_2^4
+ \psi_3 S_1^2 S_2^2
\hspace*{7mm}
\\ & &
+ \left( a^\ast a + b^\ast b \right) \left( \xi_1 S_1^2 + \xi_2 S_2^2 \right).
\ea
\es
There are seven two-particle scattering channels
($Q$ is the electric charge,
$T_3$ is the third component of weak isospin):
\begin{enumerate}
\item The channel $Q = 2,\ T_3 = 1$,
  with one state $a a$.
  \label{channel1}
\item The channel $Q = 0,\ T_3 = - 1$,
  with one state $b b$.
  \label{channel2}
\item The channel $Q = 1,\ T_3 = 0$,
  with one state $a b$.
  \label{channel3}
\item The channel $Q = 1,\ T_3 = 1$,
  with one state $a b^\ast$.
  \label{channel4}
\item The channel $Q = 1,\ T_3 = 1/2$,
  with two states $a S_1$ and $a S_2$.
  \label{channel5}
\item The channel $Q = 0,\ T_3 = -1/2$,
  with two states $b S_1$ and $b S_2$.
  \label{channel6}
\item The channel $Q = 0,\ T_3 = 0$,
  with five states $S_1^2$,
  $S_2^2$,
  $S_1 S_2$,
  $a^\ast a$,
  and $b^\ast b$.
  \label{channel7}
\end{enumerate}
In order to derive the unitarity conditions one must write the
scattering matrices for pairs of one incoming state and one outgoing state
with the same $Q$ and $T_3$.
Let the incoming state be $xy$ and let the outgoing state by $zw$,
where $x$,
$y$,
$z$,
and $w$ may be either $a$,
$a^\ast$,
$b$,
$b^\ast$,
$S_1$,
or $S_2$.
The corresponding entry in the scattering matrix
is the coefficient of $x y z^\ast w^\ast$ in $V_4$,
with the following additions:
\begin{description}
\item For each $n$ identical operators in $x y z^\ast w^\ast$
  there is an additional factor $n!$ in the entry.
\item If $x=y$ there is additional factor $2^{-1/2}$ in the entry.
\item If $z=w$ there is additional factor $2^{-1/2}$ in the entry.
\end{description}
One finds in this way that
the scattering matrices for the channels~\ref{channel1},
\ref{channel2},
\ref{channel3},
and~\ref{channel4} are
\be
\left( \begin{array}{c} \lambda_1 \end{array} \right).
\ee
The scattering matrices for the channels~\ref{channel5} and~\ref{channel6} are
\be
\left( \begin{array}{cc} 2 \xi_1 & 0 \\ 0 & 2 \xi_2 \end{array} \right).
\ee
The scattering matrix for channel~\ref{channel7} is
\be
\label{bnbj}
\left( \begin{array}{ccccc}
  6 \psi_1 & 2 \psi_3 & 0 & \sqrt{2} \xi_1 & \sqrt{2} \xi_1 \\
  2 \psi_3 & 6 \psi_2 & 0 & \sqrt{2} \xi_2 & \sqrt{2} \xi_2 \\
  0 & 0 & 4 \psi_3 & 0 & 0 \\
  \sqrt{2} \xi_1 & \sqrt{2} \xi_2 & 0 & 2 \lambda_1 & \lambda_1 \\
  \sqrt{2} \xi_1 & \sqrt{2} \xi_2 & 0 & \lambda_1 & 2 \lambda_1
\end{array} \right).
\ee
The matrix~\eqref{bnbj} is similar to the matrix
\be
\left( \begin{array}{ccccc}
  6 \psi_1 & 2 \psi_3 & 2 \xi_1 & 0 & 0 \\
  2 \psi_3 & 6 \psi_2 & 2 \xi_2 & 0 & 0 \\
  2 \xi_1 & 2 \xi_2 & 3 \lambda_1 & 0 & 0 \\
  0 & 0 & 0 & 4 \psi_3 & 0 \\
  0 & 0 & 0 & 0 & \lambda_1
\end{array} \right).
\ee

The unitarity conditions are the following:
the eigenvalues of all the scattering matrices should be smaller,
in modulus,
than $4 \pi$.
Thus,
in our case,
\bs
\label{15}
\ba
\left| \lambda_1 \right| &<& 4 \pi, \\
\left| \xi_1 \right| &<& 2 \pi, \\
\left| \xi_2 \right| &<& 2 \pi, \\
\left| \psi_3 \right| &<& \pi,
\ea
\es
and the eigenvalues of
\be
\label{16}
\left( \begin{array}{ccc}
  6 \psi_1 & 2 \psi_3 & 2 \xi_1 \\
  2 \psi_3 & 6 \psi_2 & 2 \xi_2 \\
  2 \xi_1 & 2 \xi_2 & 3 \lambda_1
\end{array} \right)
\ee
should have moduli smaller than $4 \pi$.

\subsection{Bounded-from-below conditions}

One may write
\be
\label{jhkpu}
V_4 =
\frac{1}{2}\,
\left( \begin{array}{ccc} X & Y & Z \end{array} \right)
\left( \begin{array}{ccc}
  \lambda_1 & \xi_1 & \xi_2 \\
  \xi_1 & \psi_1 & \psi_3 \\
  \xi_2 & \psi_3 & \psi_2
\end{array} \right)
\left( \begin{array}{c} X \\ Y \\ Z \end{array} \right)
\ee
where $X = \phi_1^\dagger \phi_1$,
$Y = S_1^2$,
and $Z = S_2^2$ are positive definite quantities independent of each other.
In order for $V_4$ to be positive the square matrix
  in equation~\eqref{jhkpu} must be \emph{copositive}~\cite{kannike}.
  A real symmetric matrix $M$ is copositive if $x^T M x > 0$
  for any vector $x$ with non-negative components. 
  A necessary condition for a real $n \times n$ matrix to be copositive
  is that all its $\left( n - 1 \right) \times \left( n - 1 \right)$
  principal submatrices
  are copositive too.\footnote{The principal submatrices
  are obtained by deleting rows and columns
  of the original matrix in a symmetric way,
  \textit{i.e.}~when one deletes the $i_1, i_2, \ldots, i_k$ rows
  one also deletes the $i_1, i_2, \ldots, i_k$ columns.}
  Thus,
  the matrices
  \be
  \left( \begin{array}{c} \lambda_1 \end{array} \right), \quad
  \left( \begin{array}{c} \psi_1 \end{array} \right), \quad
  \left( \begin{array}{c} \psi_2 \end{array} \right), \quad
  \left( \begin{array}{cc} \lambda_1 & \xi_1 \\ \xi_1 & \psi_1
  \end{array} \right), \quad
  \left( \begin{array}{cc} \lambda_1 & \xi_2 \\ \xi_2 & \psi_2
  \end{array} \right), \quad
  \left( \begin{array}{cc} \psi_1 & \psi_3 \\ \psi_3 & \psi_2
  \end{array} \right)
  \ee
  must be copositive.
  A real $1 \times 1$ matrix $\left( \begin{array}{c} a \end{array} \right)$
  is copositive if $a > 0$;
  a real $2 \times 2$ matrix
  $\left( \begin{array}{cc} a & c \\ c & b \end{array} \right)$
  is copositive if $a > 0$,
  $b > 0$,
  and $c > - \sqrt{a b}$.
  This leads to the six necessary BFB conditions
  \bs
  \label{BFB1}
  \ba
  \lambda_1 &>& 0, \\
  \psi_1 &>& 0, \\
  \psi_2 &>& 0, \\
  a_1 \equiv \xi_1 + \sqrt{\lambda_1 \psi_1} &>& 0, \\
  a_2 \equiv \xi_2 + \sqrt{\lambda_1 \psi_2} &>& 0, \\
  a_3 \equiv \psi_3 + \sqrt{\psi_1 \psi_2} &>& 0.
  \ea
  \es
  In order for the full $3 \times 3$ matrix
    in equation~\eqref{jhkpu}
    to be copositive an additional BFB condition is required~\cite{kannike2}:
  \be
  \label{BFB2}
  \sqrt{\lambda_1 \psi_1 \psi_2}
  + \xi_1 \sqrt{\psi_2} + \xi_2 \sqrt{\psi_1} + \psi_3 \sqrt{\lambda_1}
  + \sqrt{2 a_1 a_2 a_3} > 0.
  \ee

\subsection{Procedure}

Let the VEV of $S_1$ be $w_1$
and let the VEV of $S_2$ be $w_2$.\footnote{In
    appendix~\ref{appendixB}
    we demonstrate
    that stability points of the potential with either $w_1 = 0$ or $w_2 = 0$
    have a higher value of the potential
    and cannot therefore be the vacuum.}
Then,
the vacuum stability conditions are
\bs
\label{20}
\ba
\mu_1 &=& - \lambda_1 v^2 - \xi_1 w_1^2 - \xi_2 w_2^2, \\
m_1^2 &=& - \psi_1 w_1^2 - \psi_3 w_2^2 - \xi_1 v^2, \\
m_2^2 &=& - \psi_2 w_2^2 - \psi_3 w_1^2 - \xi_2 v^2.
\ea
\es
Using equation~\eqref{nuiho} with $G^+ = 0$ and $G^0 = 0$,
\textit{i.e.}\ in the unitary gauge,
together with $S_1 = w_1 + \sigma_1$ and $S_2 = w_2 + \sigma_2$,
one obtains
\bs
\label{cnkohp}
\ba
V &=&
- \frac{\lambda_1}{2}\, v^4
- \frac{\psi_1}{2}\, w_1^4
- \frac{\psi_2}{2}\, w_2^4
- \psi_3 w_1^2 w_2^2
- v^2 \left( \xi_1 w_1^2 + \xi_2 w_2^2 \right)
\\ & &
+ \frac{1}{2}
\left( \begin{array}{ccc} H & \sigma_1 & \sigma_2 \end{array} \right)
M
\left( \begin{array}{c} H \\ \sigma_1 \\ \sigma_2 \end{array} \right)
\label{jhuigo} \\ & &
+ \frac{\lambda_1 v}{\sqrt{2}}\, H^3
+ 2 \psi_1 w_1 \sigma_1^3
+ 2 \psi_2 w_2 \sigma_2^3
\\ & &
+ \xi_1 H \sigma_1 \left( \sqrt{2} v \sigma_1 + w_1 H \right)
+ \xi_2 H \sigma_2 \left( \sqrt{2} v \sigma_2 + w_2 H \right)
\\ & &
+ 2 \psi_3 \sigma_1 \sigma_2 \left( w_1 \sigma_2 + w_2 \sigma_1 \right)
\\ & &
+ \frac{\lambda_1}{8}\, H^4
+ \frac{\psi_1}{2}\, \sigma_1^4
+ \frac{\psi_2}{2}\, \sigma_2^4
+ \frac{\xi_1}{2}\, H^2 \sigma_1^2
+ \frac{\xi_2}{2}\, H^2 \sigma_2^2
+ \psi_3 \sigma_1^2 \sigma_2^2,
\hspace*{6mm}
\ea
\es
where
\be
\label{emghty}
M = 2 \left( \begin{array}{ccc}
  \lambda_1 v^2 &
  \sqrt{2} \xi_1 v w_1 &
  \sqrt{2} \xi_2 v w_2 \\
  \sqrt{2} \xi_1 v w_1 &
  2 \psi_1 w_1^2 &
  2 \psi_3 w_1 w_2 \\
  \sqrt{2} \xi_2 v w_2 &
  2 \psi_3 w_1 w_2 &
  2 \psi_2 w_2^2
\end{array} \right).
\ee
One diagonalizes the real symmetric matrix $M$ as
\be
\label{MMM}
M = R^T \, \mathrm{diag} \left( M_1,\, M_2,\, M_3 \right) R,
\ee
where $R$ is a $3 \times 3$ orthogonal matrix
that may be parameterized as
\be
\label{RRR}
R = \left( \begin{array}{ccc}
  c_1 & s_1 c_3 & s_1 s_3 \\
  - s_1 c_2 & c_1 c_2 c_3 + s_2 s_3 & c_1 c_2 s_3 - s_2 c_3 \\
  - s_1 s_2 & c_1 s_2 c_3 - c_2 s_3 & c_1 s_2 s_3 + c_2 c_3
\end{array} \right).
\ee
Here,
$c_j = \cos{\vartheta_j}$ and $s_j = \sin{\vartheta_j}$ for $j = 1, 2, 3$.
One has
\be
\label{RT}
\left( \begin{array}{c} H \\ \sigma_1 \\ \sigma_2 \end{array} \right)
= R^T \left( \begin{array}{c} h_1 \\ h_2 \\ h_3 \end{array} \right),
\ee
where the $h_j$ are the physical scalars,
\textit{i.e.}\ the eigenstates of mass;
the scalar $h_j$ has squared mass $M_j$.
We assume that $h_1$ is the already-observed scalar.
The interactions of the scalars with $W^+ W^-$
are given by equation~\eqref{cbuiho},
\textit{i.e.}
\be
\mathcal{L} =
\cdots + \frac{g^2 v}{\sqrt{2}}\, W_\mu^- W^{\mu +}
\left( c_1 h_1 - s_1 c_2 h_2 - s_1 s_2 h_3 \right).
\ee
We define the sign of the field $h_1$ to be such that the coupling of $h_1$
to $W^+ W^-$ has the same sign as in the Standard Model.
Thus,
we choose $- \pi/2 < \vartheta_1 < \pi/2$.

According to equation~\eqref{cnkohp},
\bs
\ba
g_3 &=& \frac{\lambda_1 v}{\sqrt{2}}\, c_1^3
+ 2 \psi_1 w_1 s_1^3 c_3^3
+ 2 \psi_2 w_2 s_1^3 s_3^3
\\ & &
+ \xi_1 c_1 s_1 c_3 \left( \sqrt{2} v s_1 c_3 + w_1 c_1 \right)
+ \xi_2 c_1 s_1 s_3 \left( \sqrt{2} v s_1 s_3 + w_2 c_1 \right)
\hspace*{7mm}
\\ & &
+ 2 \psi_3 s_1^3 c_3 s_3 \left( w_1 s_3 + w_2 c_3 \right)
\\ &=&
\frac{M_1}{2 \sqrt{2} v} \left( c_1^3
+ \frac{\sqrt{2} v}{w_1} s_1^3 c_3^3
+ \frac{\sqrt{2} v}{w_2} s_1^3 s_3^3 \right)
\\ &=&
g_3^\mathrm{SM} \left( c_1^3
+ \frac{\sqrt{2} v}{w_1} s_1^3 c_3^3
+ \frac{\sqrt{2} v}{w_2} s_1^3 s_3^3 \right),
\label{jugigo}
\ea
\es
and
\be
g_4 = \frac{\lambda_1}{8}\, c_1^4
+ \frac{\psi_1}{2}\, s_1^4 c_3^4
+ \frac{\psi_2}{2}\, s_1^4 s_3^4
+ \frac{\xi_1}{2}\, c_1^2 s_1^2 c_3^2
+ \frac{\xi_2}{2}\, c_1^2 s_1^2 s_3^2
+ \psi_3 s_1^4 c_3^2 s_3^2.
\ee
The oblique parameter $T$ is given by~\cite{vikings}
\bs
\label{t}
\ba
T = T_\mathrm{singlets} &=& \frac{3 s_1^2}{16 \pi s_w^2 m_W^2} \left\{
F \left( M_1,\, m_W^2 \right) - F \left( M_1,\, m_Z^2 \right)
\right. \\ & &
- c_2^2
\left[ F \left( M_2,\, m_W^2 \right) - F \left( M_2,\, m_Z^2 \right) \right]
\\ & & \left.
- s_2^2
\left[ F \left( M_3,\, m_W^2 \right) - F \left( M_3,\, m_Z^2 \right) \right]
\right\},
\ea
\es
where
\be
F \left( x,\, y \right) = \left\{
\begin{array}{lcl}
  {\displaystyle \frac{x + y}{2} - \frac{x y}{x - y}\, \ln{\frac{x}{y}}} &
      \Leftarrow & x \neq y, \\
  0 & \Leftarrow & x = y.
\end{array} \right.
\ee

In our numerical work we use as input the nine quantities $v$,
$w_1$,
$w_2$,
$M_1$,
$M_2$,
$M_3$,
$\vartheta_1$,
$\vartheta_2$,
and $\vartheta_3$,
which are equivalent to the nine parameters of the scalar potential $\mu_1$,
$m_1^2$,
$m_2^2$,
$\lambda_1$,
$\psi_1$,
$\psi_2$,
$\psi_3$,
$\xi_1$,
and $\xi_2$.
We input equations~\eqref{data}
and choose arbitrary values for $M_2 > 0$ and $M_3 > 0$ such that $M_2 \le M_3$
(this represents no lack of generality,
it is just the naming convention for $h_2$ and $h_3$).
We enforce no lower bound on $M_2$ and $M_3$,
  in particular we allow them to be lower
  than $M_1 = \left( 125\,\mathrm{GeV} \right)^2$.
The VEVs $w_1$ and $w_2$ are chosen positive;
this corresponds to the freedom of choice of the signs of $S_1$ and $S_2$.
The angle $\vartheta_1$ is in either the first or the fourth quadrant,
with
\be
\label{c1}
\cos{\vartheta_1} > 0.9,
\ee
so that the $h_1 W^+ W^-$ coupling is within 10\%
of its Standard Model value.
The angle $\vartheta_2$ is in the first quadrant;
this corresponds to a choice of the signs of the fields $h_2$ and $h_3$.
The angle $\vartheta_3$ may be in any quadrant.
We firstly compute $T$ according to equation~\eqref{t}
and check that it is inside its experimentally allowed domain~\cite{RPP}
$-0.04 < T < 0.20$.
We then compute
\bs
\ba
\lambda_1 &=& \frac{1}{2 v^2} \left(
M_1 c_1^2 + M_2 s_1^2 c_2^2 + M_3 s_1^2 s_2^2 \right),
\\
\psi_1 &=& \frac{1}{4 w_1^2} \left[ M_1 s_1^2 c_3^2
  + M_2 \left( c_1 c_2 c_3 + s_2 s_3 \right)^2
  + M_3 \left( c_1 s_2 c_3 - c_2 s_3 \right)^2 \right],
\\
\psi_2 &=& \frac{1}{4 w_2^2} \left[ M_1 s_1^2 s_3^2
  + M_2 \left( c_1 c_2 s_3 - s_2 c_3 \right)^2
  + M_3 \left( c_1 s_2 s_3 + c_2 c_3 \right)^2 \right],
\\
\xi_1 &=& \frac{1}{2 \sqrt{2} v w_1} \left[ M_1 c_1 s_1 c_3
  - M_2 c_1 s_1 c_2^2 c_3 - M_3 c_1 s_1 s_2^2 c_3
  \right. \no & & \left.
  + \left( M_3 - M_2 \right) s_1 c_2 s_2 s_3 \right],
\\
\xi_2 &=& \frac{1}{2 \sqrt{2} v w_2} \left[ M_1 c_1 s_1 s_3
  - M_2 c_1 s_1 c_2^2 s_3 - M_3 c_1 s_1 s_2^2 s_3
  \right. \no & & \left.
  + \left( M_2 - M_3 \right) s_1 c_2 s_2 c_3 \right],
\\
\psi_3 &=& \frac{1}{4 w_1 w_2} \left[ M_1 s_1^2 c_3 s_3
  + M_2 \left( c_1^2 c_2^2 - s_2^2 \right) c_3 s_3
  + M_3 \left( c_1^2 s_2^2 - c_2^2 \right) c_3 s_3
  \right. \no & & \left.
  + \left( M_3 - M_2 \right) c_1 c_2 s_2 \left( c_3^2 - s_3^2 \right) \right].
\ea
\es
We validate the input if the
inequalities~\eqref{15},
  \eqref{BFB1},
  and~\eqref{BFB2}
hold
and if the moduli of all three eigenvalues of the matrix~\eqref{16}
are smaller than $4 \pi$.

\subsection{Results}

\begin{figure}
\begin{center}
\epsfig{file=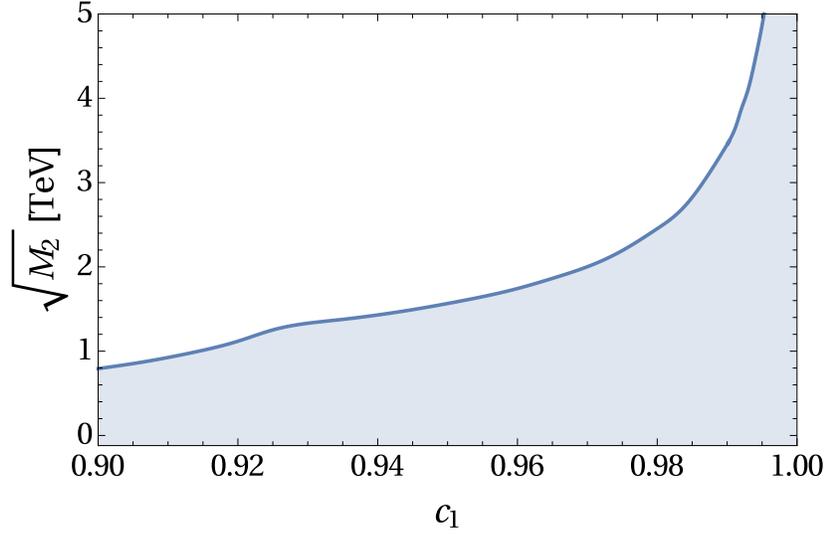,width=0.7\textwidth}
\end{center}
\caption{The upper bound on the mass of the second-lightest scalar $\sqrt{M_2}$
  \textit{versus} $\cos{\vartheta_1}$ in the SM2S. \label{fig01}}
\end{figure}

A remarkable result of our numerical work is that there
is an upper bound on the mass $\sqrt{M_2}$;
even if the VEVs $w_1$ and $w_2$ are allowed to be as high as 100\,TeV---and,
correspondingly,
the mass $\sqrt{M_3}$ also grows to a value of that order---the mass
$\sqrt{M_2}$ remains much smaller.
In figure~\ref{fig01} we depict the upper bound on $\sqrt{M_2}$
as a function of $c_1$;
when $c_1 \to 1$ the upper bound disappears,
\textit{i.e.}\ it tends to infinity.
  We emphasize that the bound depicted by the solid line
  in figure~\ref{fig01} was obtained through a random scan
  of the parameter space;
  it is not an analytical bound.

\begin{figure}
\begin{center}
\epsfig{file=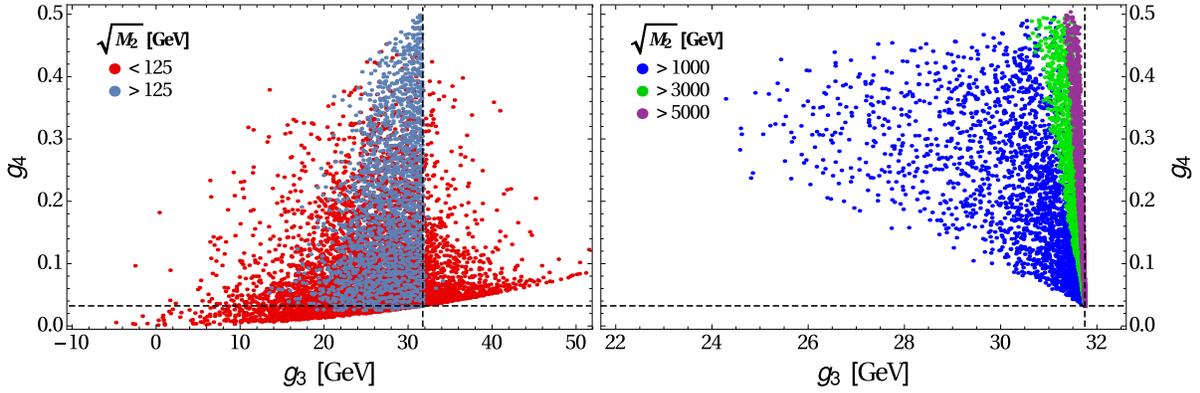,width=1.0\textwidth}
\end{center}
\caption{Scatter plots of the four-Higgs coupling $g_4$ \textit{versus}\/
  the three-Higgs coupling $g_3$ in the SM2S.
  This figure was produced by randomly generating
    $\sqrt{M_2},\, \sqrt{M_3},\, w_1,\, w_2
    \in \left[ 0\,\mathrm{TeV},\ 10\,\mathrm{TeV} \right]$.
  The dashed lines mark the values of $g_3$ and $g_4$ in the SM.
  The left panel includes both red points with $M_2 < M_1$
  and grey points with $M_2 > M_1$;
  the right panel depicts points
  that have $\sqrt{M_2}$ larger than either 1, 3, or 5\,TeV.
  (In order not to overcrowd the left panel,
  we have used in it just a subset
  of the set of large-$M_2$ points that we have generated.)
  \label{fig02}}
\end{figure}

In figure~\ref{fig02} we display the predictions for $g_3$ and $g_4$.
  In order to produce that figure
  we have randomly generated $\sqrt{M_2}$,
  $\sqrt{M_3}$,
  and the VEVs $w_1$ and $w_2$ in the range 0 to 10~TeV.
  One sees that $g_3$ is always positive but below its SM value
  when $M_2 > M_1$;
  when $M_2 < M_1$ the allowed range for $g_3$ becomes much wider.
  When the masses of the new scalars get higher,
  $g_3$ takes values closer to the SM value.
  An important point is that $g_3$ remains of the same order of magnitude
  as in the SM,
  but $g_4$ may reach 15 times its SM value.

\begin{figure}
\begin{center}
\epsfig{file=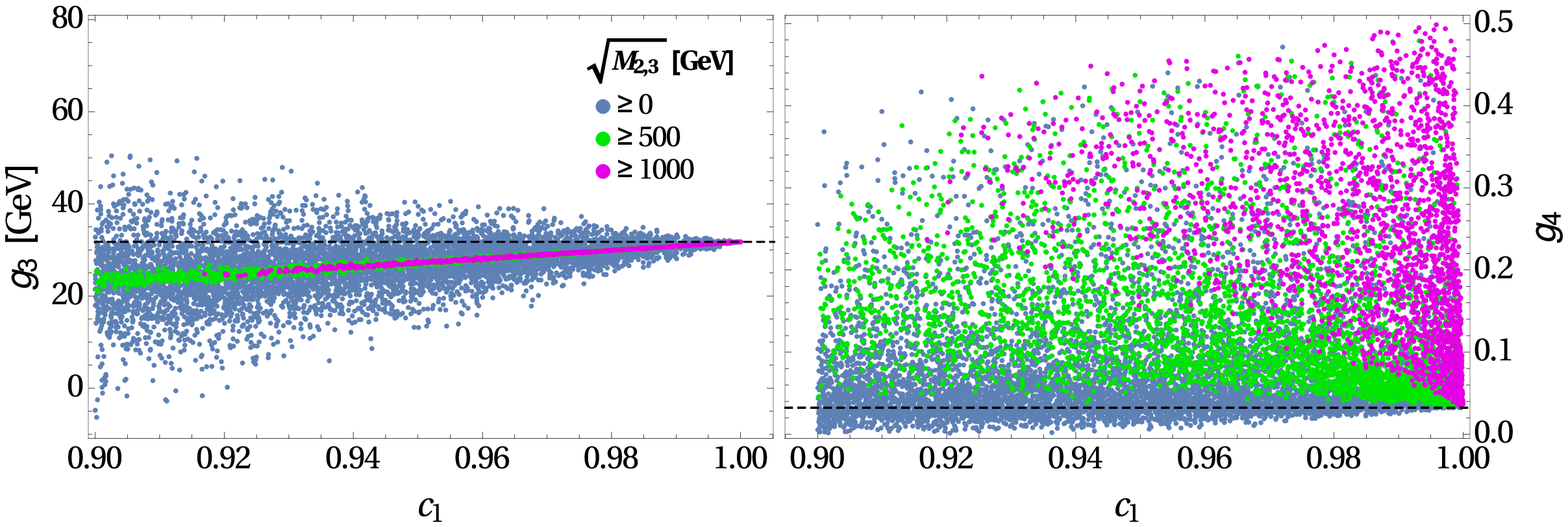,width=1.0\textwidth}
\end{center}
\caption{Scatter plots of the three-Higgs coupling $g_3$
  (left panel)
  and of the four-Higgs coupling $g_4$
  (right panel)
  \textit{versus}\/ $\cos{\vartheta_1}$ in the SM2S.
  The dashed lines mark the values of the couplings in the SM.
  \label{fig03}}
\end{figure}

In the left panel of figure~\ref{fig03}
one sees that
  when $\cos{\vartheta_1} \to 1$ the coupling $g_3$
  necessarily approaches its SM value.
  This behaviour is because of equation~\eqref{jugigo} and $c_1 > 0.9$,
  which implies $\left| s_1 \right| \ll c_1$.
  On the other hand,
  $g_4$ is not correlated with $\cos{\vartheta_1}$,
  as one sees in the right panel of figure~\ref{fig03}.

\section{The two-Higgs-doublet model} \label{sec:2HDM}

We next consider the model with two scalar gauge-$SU(2)$ doublets
$\phi_1$ and $\phi_2$ having the same weak hypercharge.
This is usually known as 2HDM.
The scalar potential is given by equation~\eqref{hoihp},
where
\bs
\label{gjihoree}
\ba
V_2 &=& \mu_1 \phi_1^\dagger \phi_1 + \mu_2 \phi_2^\dagger \phi_2
+ \left( \mu_3 \phi_1^\dagger \phi_2 + \mathrm{H.c.} \right),
\label{gmkho} \\
V_4 &=&
\frac{\lambda_1}{2} \left( \phi_1^\dagger \phi_1 \right)^2
+ \frac{\lambda_2}{2} \left( \phi_2^\dagger \phi_2 \right)^2
+ \lambda_3\, \phi_1^\dagger \phi_1\, \phi_2^\dagger \phi_2
+ \lambda_4\, \phi_1^\dagger \phi_2\, \phi_2^\dagger \phi_1
\hspace*{7mm} \\ & &
+ \left[
  \frac{\lambda_5}{2} \left( \phi_1^\dagger \phi_2 \right)^2
  + \lambda_6\, \phi_1^\dagger \phi_1\, \phi_1^\dagger \phi_2
  + \lambda_7\, \phi_2^\dagger \phi_2\, \phi_1^\dagger \phi_2
  + \mathrm{H.c.}
  \right],
\ea
\es
where $\mu_{1,2}$ and $\lambda_{1,2,3,4}$ are real.
The ten (real) coefficients in $V_4$ may be grouped as~\cite{maniatis}
\bs
\ba
\eta_{00} &=& \lambda_1 + \lambda_2 + 2 \lambda_3, \\*[3mm]
\eta = \left( \begin{array}{c} \eta_1 \\ \eta_2 \\ \eta_3 \end{array} \right)
&=& \left( \begin{array}{c}
  2\, \Re{\left( \lambda_6 + \lambda_7 \right)} \\
  - 2\, \Im{\left( \lambda_6 + \lambda_7 \right)} \\
  \lambda_1 - \lambda_2
\end{array} \right),
\\*[3mm]
E =
\left( \begin{array}{ccc} \eta_{11} & \eta_{12} & \eta_{13} \\
  \eta_{12} & \eta_{22} & \eta_{23} \\ \eta_{13} & \eta_{23} & \eta_{33}
\end{array} \right)
&=&
\left( \begin{array}{ccc}
  2 \lambda_4 + 2\, \Re{\lambda_5} &
  - 2\, \Im{\lambda_5} &
  2\, \Re{\left( \lambda_6 - \lambda_7 \right)} \\
  - 2\, \Im{\lambda_5} &
  2 \lambda_4 - 2\, \Re{\lambda_5} &
  - 2\, \Im{\left( \lambda_6 - \lambda_7 \right)} \\
  2\, \Re{\left( \lambda_6 - \lambda_7 \right)} &
  - 2\, \Im{\left( \lambda_6 - \lambda_7 \right)} &
  \lambda_1 + \lambda_2 - 2 \lambda_3
\end{array} \right). \hspace*{6mm}
\ea
\es
Under a (unitary) change of basis of the scalar doublets,
$\eta_{00}$ is invariant while
\be
\label{uighu}
\eta \to O \eta, \quad E \to O E O^T,
\ee
where $O$ is an $SO(3)$ matrix.
Only quantities and procedures
that are invariant under the transformation~\eqref{uighu} are meaningful.

\subsection{Unitarity conditions} \label{uhi}

We write
\be
\label{mblghph}
\phi_1 = \left( \begin{array}{c} a \\ b \end{array} \right),
\quad
\phi_2 = \left( \begin{array}{c} c \\ d \end{array} \right),
\quad
\phi_1^\dagger = \left( \begin{array}{cc} a^\ast & b^\ast \end{array} \right),
\quad
\phi_2^\dagger = \left( \begin{array}{cc} c^\ast & d^\ast \end{array} \right).
\ee
Then,
\bs
\ba
V_4 &=&
\lambda_1 \left( \frac{a^\ast a^\ast a a + b^\ast b^\ast b b}{2}
+ a^\ast b^\ast a b \right)
\\ & &
+ \lambda_2 \left( \frac{c^\ast c^\ast c c + d^\ast d^\ast d d}{2}
+ c^\ast d^\ast c d \right)
\\ & &
+ \left( \lambda_3 + \lambda_4 \right)
\left( a^\ast c^\ast a c + b^\ast d^\ast b d \right)
\\ & &
+ \lambda_3 \left( a^\ast d^\ast a d + b^\ast c^\ast b c \right)
\\ & &
+ \lambda_4 \left( a^\ast d^\ast b c + b^\ast c^\ast a d \right)
\\ & &
+ \lambda_5 \left( \frac{a^\ast a^\ast c c + b^\ast b^\ast d d}{2}
+ a^\ast b^\ast c d \right)
\\ & &
+ \lambda_5^\ast \left( \frac{c^\ast c^\ast a a + d^\ast d^\ast b b}{2}
+ c^\ast d^\ast a b \right)
\\ & &
+ \lambda_6 \left( a^\ast a^\ast a c + b^\ast b^\ast b d
+ a^\ast b^\ast a d + a^\ast b^\ast b c \right)
\\ & &
+ \lambda_6^\ast \left( a^\ast c^\ast a a + b^\ast d^\ast b b
+ a^\ast d^\ast a b + b^\ast c^\ast a b \right)
\\ & &
+ \lambda_7 \left( a^\ast c^\ast c c + b^\ast d^\ast d d
+ b^\ast c^\ast c d + a^\ast d^\ast c d \right)
\\ & &
+ \lambda_7^\ast \left( c^\ast c^\ast a c + d^\ast d^\ast b d
+ c^\ast d^\ast b c + c^\ast d^\ast a d \right).
\ea
\es
The relevant scattering channels are~\cite{silva}:
\begin{enumerate}
\item The channel $Q = 2,\ T_3 = 1$,
  with three states $aa$,
  $cc$,
  and $ac$.
  \label{channel10}
\item The channel $Q = 0,\ T_3 = -1$,
  with three states $bb$,
  $dd$,
  and $bd$.
  \label{channel20}
\item The channel $Q=1,\ T_3 = 0$,
  with four states $ab$,
  $cd$,
  $ad$,
  and $bc$.
  \label{channel30}
\item The channel $Q = 1,\ T_3 = 1$,
  with four states $a b^\ast$,
  $c d^\ast$,
  $a d^\ast$,
  and $c b^\ast$.
  \label{channel40}
\item The channel $Q = 0,\ T_3 = 0$,
  with eight states $aa^\ast$,
  $bb^\ast$,
  $cc^\ast$,
  $dd^\ast$,
  $ac^\ast$,
  $bd^\ast$,
  $ca^\ast$,
  and $db^\ast$.
  \label{channel50}
\end{enumerate}
Channel~\ref{channel50} produces the scattering matrix
\be
\label{vidkgf}
\left( \begin{array}{cccccccc}
  2 \lambda_1 & \lambda_1 & \lambda_3 + \lambda_4 & \lambda_3 &
  2 \lambda_6^\ast & \lambda_6^\ast & 2 \lambda_6 & \lambda_6 \\
  \lambda_1 & 2 \lambda_1 & \lambda_3 & \lambda_3 + \lambda_4 &
  \lambda_6^\ast & 2 \lambda_6^\ast & \lambda_6 & 2 \lambda_6 \\
  \lambda_3 + \lambda_4 & \lambda_3 & 2 \lambda_2 & \lambda_2 &
  2 \lambda_7^\ast & \lambda_7^\ast & 2 \lambda_7 & \lambda_7 \\
  \lambda_3 & \lambda_3 + \lambda_4 & \lambda_2 & 2 \lambda_2 &
  \lambda_7^\ast & 2 \lambda_7^\ast & \lambda_7 & 2 \lambda_7 \\
  2 \lambda_6 & \lambda_6 & 2 \lambda_7 & \lambda_7 &
  \lambda_3 + \lambda_4 & \lambda_4 & 2 \lambda_5 & \lambda_5 \\
  \lambda_6 & 2 \lambda_6 & \lambda_7 & 2 \lambda_7 &
  \lambda_4 & \lambda_3 + \lambda_4 & \lambda_5 & 2 \lambda_5 \\
  2 \lambda_6^\ast & \lambda_6^\ast & 2 \lambda_7^\ast & \lambda_7^\ast &
  2 \lambda_5^\ast & \lambda_5^\ast & \lambda_3 + \lambda_4 & \lambda_4 \\
  \lambda_6^\ast & 2 \lambda_6^\ast & \lambda_7^\ast & 2 \lambda_7^\ast &
  \lambda_5^\ast & 2 \lambda_5^\ast & \lambda_4 & \lambda_3 + \lambda_4
\end{array} \right).
\ee
A similarity transformation transforms the matrix~\eqref{vidkgf} into
the direct sum of two $4 \times 4$ matrices
\bs
\label{dygut}
\ba
\mathcal{M}_1 &=& \frac{1}{2} \left( \begin{array}{cc}
  \eta_{00} - 2 I & \eta^T \\ \eta & E + 2 I \times \mathbbm{1}_{3 \times 3}
\end{array} \right),
\label{jhui} \\*[3mm]
\mathcal{M}_2 &=& \frac{1}{2} \left( \begin{array}{cc}
  3 \eta_{00} - 2 I & 3 \eta^T \\
  3 \eta & 3 E + 2 I \times \mathbbm{1}_{3 \times 3}
\end{array} \right).
\label{uhiooo}
\ea
\es
Here,
\be
\label{I}
I = \lambda_3 - \lambda_4 = \frac{\eta_{00} - \mathrm{tr}\, E}{4}
\ee
is invariant under a change of basis of the doublets.
It is obvious that the eigenvalues of the matrices~\eqref{dygut}
are invariant under such a change too.

Channel~\eqref{channel40} produces the scattering matrix
\be
\left( \begin{array}{cccc}
  \lambda_1 & \lambda_4 & \lambda_6^\ast & \lambda_6 \\
  \lambda_4 & \lambda_2 & \lambda_7^\ast & \lambda_7 \\
  \lambda_6 & \lambda_7 & \lambda_3 & \lambda_5 \\
  \lambda_6^\ast & \lambda_7^\ast & \lambda_5^\ast & \lambda_3
\end{array} \right),
\ee
which may readily be shown to be similar to $\mathcal{M}_1$.
Channel~\eqref{channel30} produces the scattering matrix
\be
\label{ugihop}
\left( \begin{array}{cccc}
  \lambda_1 & \lambda_5 & \lambda_6 & \lambda_6 \\
  \lambda_5^\ast & \lambda_2 & \lambda_7^\ast & \lambda_7^\ast \\
  \lambda_6^\ast & \lambda_7 & \lambda_3 & \lambda_4 \\
  \lambda_6^\ast & \lambda_7 & \lambda_4 & \lambda_3
\end{array} \right).
\ee
The matrix~\eqref{ugihop} is similar to
\be
\left( \begin{array}{cccc}
  & & & 0 \\
  & \mathcal{M}_3 & & 0 \\
  & & & 0 \\
  0 & 0 & 0 & I
\end{array} \right),
\ee
where
\be
\label{buigpj}
\mathcal{M}_3 = \left( \begin{array}{ccc}
  \lambda_1 & \lambda_5 & \sqrt{2} \lambda_6 \\
  \lambda_5^\ast & \lambda_2 & \sqrt{2} \lambda_7^\ast \\
  \sqrt{2} \lambda_6^\ast & \sqrt{2} \lambda_7 & \lambda_3 + \lambda_4
\end{array} \right).
\ee
Channels~\eqref{channel10} and~\eqref{channel20}
also lead to the matrix $\mathcal{M}_3$.
Direct computation demonstrates that the eigenvalues of $\mathcal{M}_3$
are invariant under the transformation~\eqref{uighu}.

Thus,
the unitarity conditions for the scalar potential of the 2HDM
are the following:
the eigenvalues of the two $4 \times 4$ matrices~\eqref{dygut}
and of the $3 \times 3$ matrix~\eqref{buigpj},
and $I$ in equation~\eqref{I},
should have moduli smaller than $4 \pi$.
These conditions were first derived in ref.~\cite{theJapanese}.
We emphasize that they are,
as they should,
invariant under a change of basis of the two doublets.

\subsubsection{The case $\lambda_6 = \lambda_7 = 0$}

If $\lambda_6 = \lambda_7 = 0$,
then $\eta_1 = \eta_2 = \eta_{13} = \eta_{23} = 0$
and this simplifies things considerably.
The unitarity conditions are then
\bs
\label{hgjkfj}
\ba
\left| \lambda_3 + \lambda_4 \right| &<& 4 \pi, \label{a1} \\
\left| \lambda_3 - \lambda_4 \right| &<& 4 \pi, \label{a2} \\
\left| \lambda_3 + \left| \lambda_5 \right| \right| &<& 4 \pi, \label{a3} \\
\left| \lambda_3 - \left| \lambda_5 \right| \right| &<& 4 \pi, \label{a4} \\
a_+ \equiv \left| \lambda_3 + 2 \lambda_4
+ 3 \left| \lambda_5 \right| \right| &<& 4 \pi, \label{a5} \\
a_- \equiv \left| \lambda_3 + 2 \lambda_4
- 3 \left| \lambda_5 \right| \right| &<& 4 \pi, \label{a6} \\
\left| \lambda_1 + \lambda_2 + \sqrt{\left( \lambda_1 - \lambda_2 \right)^2
  + 4 \left| \lambda_5 \right|^2} \right| &<& 8 \pi, \label{a7} \\
\left| \lambda_1 + \lambda_2 - \sqrt{\left( \lambda_1 - \lambda_2 \right)^2
  + 4 \left| \lambda_5 \right|^2} \right| &<& 8 \pi, \label{a8} \\
\left| \lambda_1 + \lambda_2 + \sqrt{\left( \lambda_1 - \lambda_2 \right)^2
  + 4 \lambda_4^2} \right| &<& 8 \pi, \label{a9} \\
\left| \lambda_1 + \lambda_2 - \sqrt{\left( \lambda_1 - \lambda_2 \right)^2
  + 4 \lambda_4^2} \right| &<& 8 \pi, \label{a10} \\
b_+ \equiv \left| 3 \lambda_1 + 3 \lambda_2
+ \sqrt{9 \left( \lambda_1 - \lambda_2 \right)^2
  + 4 \left( 2 \lambda_3 + \lambda_4 \right)^2} \right| &<& 8 \pi, \label{a11}
\\
b_- \equiv \left| 3 \lambda_1 + 3 \lambda_2
- \sqrt{9 \left( \lambda_1 - \lambda_2 \right)^2
  + 4 \left( 2 \lambda_3 + \lambda_4 \right)^2} \right| &<& 8 \pi. \label{a12}
\ea
\es

\subsubsection{The case $\lambda_1 = \lambda_2
  = \lambda_3 = \lambda_4 = \lambda_5 = 0$}

The case $\lambda_1 = \lambda_2
= \lambda_3 = \lambda_4 = \lambda_5 = 0$ is not realistic
because it produces a potential unbounded from below.
Still,
one may compute the unitarity conditions in that case and one obtains
\bs
\label{bjihug}
\ba
\sqrt{\left| \lambda_6 \right|^2 + \left| \lambda_7 \right|^2}
&<& 2 \sqrt{2} \pi, \label{b1} \\
\sqrt{\left| \lambda_6 \right|^2 + \left| \lambda_7 \right|^2
  + \left| \lambda_6^2 + \lambda_7^2 \right|}
&<& \frac{4 \pi}{3}. \label{b2}
\ea
\es

\subsubsection{Consequences}

\begin{figure}
\begin{center}
\epsfig{file=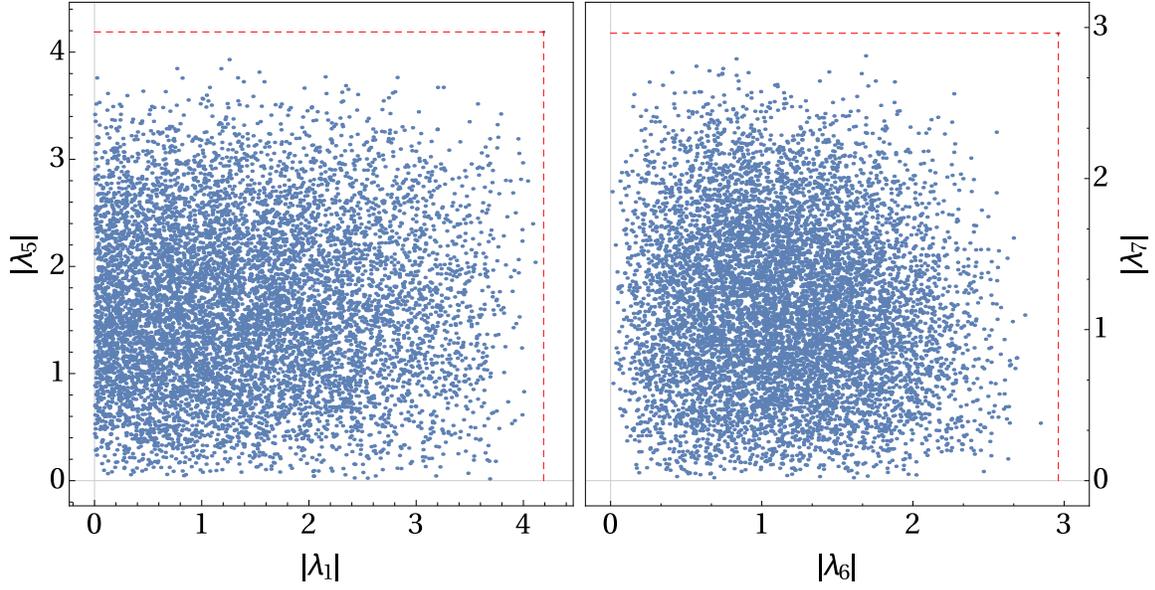,width=1.0\textwidth}
\end{center}
\caption{Scatter plots of $\left| \lambda_1 \right|$
  \textit{versus} $\left| \lambda_5 \right|$
  and of $\left| \lambda_6 \right|$
  \textit{versus} $\left| \lambda_7 \right|$
  with the unitarity conditions enforced.
  The dashed red lines indicate the bounds $\left| \lambda_{1,5} \right|
  < 4 \pi / 3$ and $\left| \lambda_{6,7} \right| < 2 \sqrt{2} \pi / 3$,
  respectively.
  \label{fig10}}
\end{figure}

\begin{figure}
\begin{center}
\epsfig{file=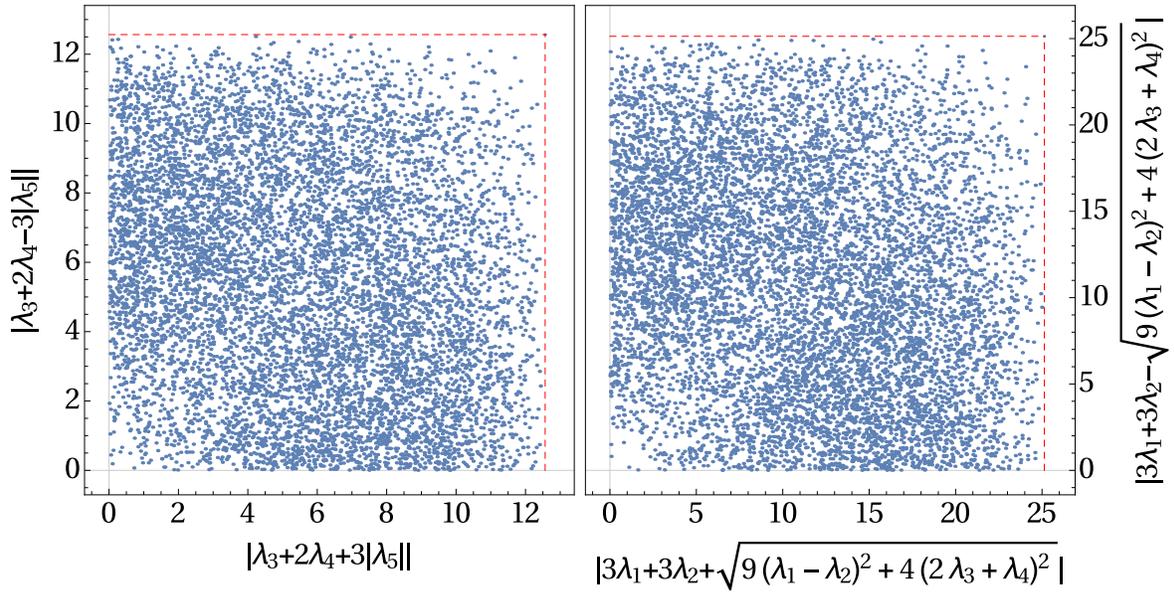,width=1.0\textwidth}
\end{center}
\caption{Scatter plots of $a_\pm$ and $b_\pm$---see
  equations~\eqref{a5},
  \eqref{a6},
  \eqref{a11},
  and~\eqref{a12}---with the unitarity conditions enforced.
  The red dashed lines indicate the bounds
  $a_\pm < 4 \pi$ in the left plot and $b_\pm < 8 \pi$ in the right plot.
  \label{fig11}}
\end{figure}

\begin{figure}
\begin{center}
\epsfig{file=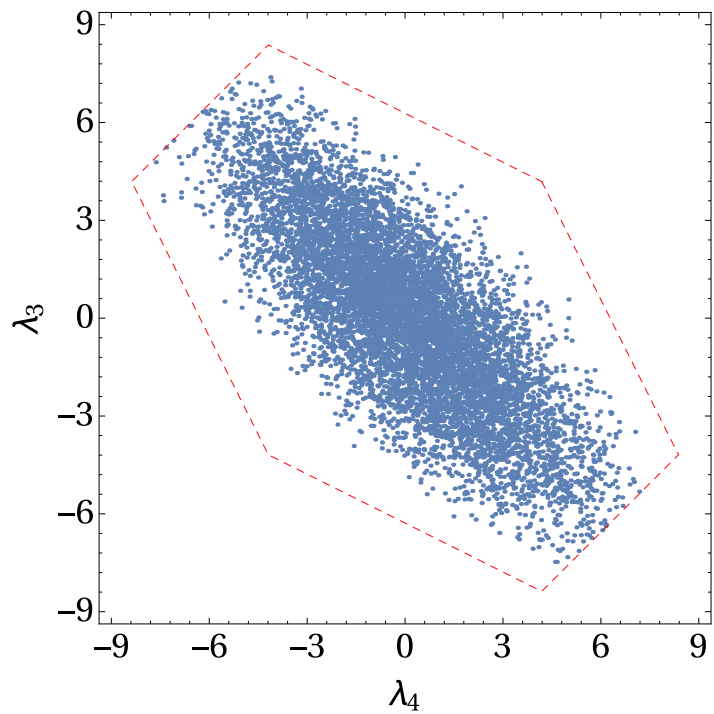,width=0.6\textwidth}
\end{center}
\caption{Scatter plots of $\lambda_3$ \textit{versus} $\lambda_4$
  with the unitarity conditions enforced.
  The dashed red lines are given by the equations
  $\left| \lambda_3 - \lambda_4 \right| = 4 \pi$,
  $\left| 2 \lambda_3 + \lambda_4 \right| = 4 \pi$,
  and $\left| \lambda_3 + 2 \lambda_4 \right| = 4 \pi$.
  \label{fig12}}
\end{figure}

We have numerically analyzed the unitarity conditions
by giving random values to $\lambda_1$,
$\lambda_2$,
$\lambda_3$,
$\lambda_4$,
$\left| \lambda_5 \right|$,
$\left| \lambda_6 \right|$,
$\left| \lambda_7 \right|$,
$\arg{\left( \lambda_5^\ast \lambda_6 \lambda_7 \right)}$,
and $\arg{\left( \lambda_6^\ast \lambda_7 \right)}$ and then checking
whether all the unitarity conditions are met.
We present in figures~\ref{fig10}--\ref{fig12}
scatter plots with more than 8,000 allowed points each.
We have found that
all the conditions~\eqref{hgjkfj}
still hold even when $\lambda_6 = \lambda_7 = 0$ is not true;
also,
the conditions~\eqref{bjihug} still hold even when
$\lambda_1 = \lambda_2 = \lambda_3 = \lambda_4 = \lambda_5 = 0$ does not apply.
In particular,
the upper bounds~\eqref{a2},
\eqref{a5},
\eqref{a6},
\eqref{a11},
and~\eqref{a12} are sometimes attained,
as illustrated in figures~\ref{fig12} and~\ref{fig11},
respectively.
For the individual parameters,
the bounds
\bs
\ba
\left| \lambda_{1,2} \right| &<& \frac{4 \pi}{3}, \label{huigoy} \\
\left| \lambda_5 \right| &<& \frac{4 \pi}{3}, \label{huigoy2} \\
\left| \lambda_{6,7} \right| &<& \frac{2 \sqrt{2} \pi}{3} \label{huigoy3}
\ea
\es
hold and are illustrated in figure~\ref{fig10};
the bound~\eqref{huigoy} is suggested by inequality~\eqref{a11}
when $\lambda_3$,
$\lambda_4$,
and either $\lambda_1$ or $\lambda_2$ vanish;
the bound~\eqref{huigoy2} is suggested by inequality~\eqref{a5}
when $\lambda_3 = \lambda_4 = 0$,
and the bound~\eqref{huigoy3} is suggested by inequality~\eqref{b2}
when either $\lambda_6$ or $\lambda_7$ vanishes.
Finally,
$\left( \lambda_3,\, \lambda_4 \right)$ is always within the hexagon
with sides $\left| \lambda_3 - \lambda_4 \right| = 4 \pi$,
$\left| 2 \lambda_3 + \lambda_4 \right| = 4 \pi$,
and $\left| \lambda_3 + 2 \lambda_4 \right| = 4 \pi$,
as illustrated in figure~\ref{fig12}.

\subsection{Bounded-from-below conditions} \label{uhi2}

Necessary and sufficient conditions for the scalar potential of the 2HDM
to be BFB were first derived in ref.~\cite{maniatis}.
Ivanov~\cite{ivanov} and Silva~\cite{silva-ivanov}
later produced other,
equivalent conditions to the same effect.
We have implemented numerically both the conditions of ref.~\cite{maniatis}
and those of ref.~\cite{silva-ivanov}.
We have found that the Ivanov--Silva algorithm runs several times faster
than the one of ref.~\cite{maniatis}.
We have also checked that all the points produced by either algorithm
were validated by the other one.

The points in our scatter plots were produced by using the algorithm
of ref.~\cite{silva-ivanov}.
That algorithm runs as follows.
One constructs the $4 \times 4$ matrix
\be
\label{lambdae}
\Lambda_E = \left( \begin{array}{cc} \eta_{00} & \eta^T \\ - \eta & - E
\end{array} \right)
\ee
and one computes its four eigenvalues.
Then the potential is BFB
if all the following conditions apply:
\begin{itemize}
\item All four eigenvalues are real.
\item All four eigenvalues are different from each other.
\item Call $\Lambda_0$ the largest eigenvalue.
  Call the other three eigenvalues $\Lambda_{1,2,3}$.
  The eigenvalue $\Lambda_0$ is positive;
  thus,
  \be
  \Lambda_0 > \Lambda_{1,2,3}, \quad \Lambda_0 > 0.
  \ee
  (Each of $\Lambda_1$,
  $\Lambda_2$,
  and $\Lambda_3$ may be either positive or negative.)
\item
  \be
  \label{gohpu}
  \frac{\left[ \left( \Lambda_E - \Lambda_1 \times \mathbbm{1}_{4 \times 4}
      \right)
    \times \left( \Lambda_E - \Lambda_2 \times \mathbbm{1}_{4 \times 4} \right)
    \times \left( \Lambda_E - \Lambda_3 \times \mathbbm{1}_{4 \times 4} \right)
    \right]_{11}}{\left( \Lambda_0 - \Lambda_1 \right)
    \left( \Lambda_0 - \Lambda_2 \right) \left( \Lambda_0 - \Lambda_3 \right)}
  > 0.
  \ee
\end{itemize}

It is possible to derive analytically some necessary conditions
for boundedness-from-below.
Let us parameterize
\be
\label{r2}
\phi_1^\dagger \phi_1 = r^2 \sin^2{\theta}, \quad
\phi_2^\dagger \phi_2 = r^2 \cos^2{\theta}, \quad
\phi_1^\dagger \phi_2 = e^{i \alpha} r^2 h \sin{\theta} \cos{\theta},
\ee
where $0 \le \theta \le \pi/2$ without loss of generality.
Since,
in the notation of equations~\eqref{mblghph},
\be
r^4 \left( 1 - h^2 \right) \sin^2{\theta} \cos^2{\theta}
= \phi_1^\dagger \phi_1\, \phi_2^\dagger \phi_2
- \phi_1^\dagger \phi_2\, \phi_2^\dagger \phi_1
= \left| a d - b c \right|^2 \ge 0,
\ee
one concludes that $h^2 \le 1$.
Thus,
without loss of generality
$0 \le h \le 1$ while the phase $\alpha$ is arbitrary.
Boundedness from below of $V_4$ means that
\bs
\label{mnsdfy}
\ba
\frac{\lambda_1}{2} \sin^4{\theta}
+ \frac{\lambda_2}{2} \cos^4{\theta}
+ \left[ \lambda_3 + \lambda_4 h^2
  + \Re{\left( \lambda_5 e^{2 i \alpha} \right)} h^2 \right]
\sin^2{\theta} \cos^2{\theta}
& & \\
+ 2 h\, \Re{\left( \lambda_6 e^{i \alpha} \right)} \sin^3{\theta} \cos{\theta}
+ 2 h\, \Re{\left( \lambda_7 e^{i \alpha} \right)} \sin{\theta} \cos^3{\theta}
&>& 0
\ea
\es
for any $\theta$,
$h$,
and $\alpha$. From the cases $\theta = 0$ and $\theta = \pi/2$
one derives
\be
\label{huihu}
\lambda_1 > 0, \quad \lambda_2 > 0.
\ee
Making $\alpha \to \pi + \alpha$ in inequality~\eqref{mnsdfy},
one concludes that
\bs
\label{bjghiu}
\ba
2 h \sin{\theta} \cos{\theta}
\left| \Re{\left[ \left( \lambda_6 \sin^2{\theta}
    + \lambda_7 \cos^2{\theta} \right) e^{i \alpha} \right]} \right|
&<& 
\frac{\lambda_1}{2} \sin^4{\theta} + \frac{\lambda_2}{2} \cos^4{\theta}
\\ & &
+ \left[ \lambda_3 + \lambda_4 h^2
  \right. \\ & & \left.
  + \Re{\left( \lambda_5 e^{2 i \alpha} \right)} h^2 \right]
\sin^2{\theta} \cos^2{\theta}.
\ea
\es
Therefore,
the quantity in the right-hand side of inequality~\eqref{bjghiu}
must be positive for any $\theta$,
$h$,
and $\alpha$.
It is easy to see that
\be
\label{fmbjkho}
\varrho \sin^4{\theta} + \varsigma \cos^4{\theta}
+ \varepsilon \sin^2{\theta} \cos^2{\theta} > 0
\ \, \forall \theta \in \left[ 0,\ \frac{\pi}{2} \right]
\quad \Leftrightarrow \quad
\varrho > 0,\ \varsigma > 0,\ \varepsilon > -2 \sqrt{\varrho \varsigma}.
\ee
Applying the statement~\eqref{fmbjkho}
to the case $\varrho = \lambda_1 / 2$,
$\varsigma = \lambda_2 / 2$,
$\varepsilon = \lambda_3 + \lambda_4 h^2
+ \Re{\left( \lambda_5 e^{2 i \alpha} \right)} h^2$
for any $h \in \left[ 0,\ 1 \right]$ and $\alpha$,
one concludes that
\be
\label{mnjoh}
\lambda_3 > - \sqrt{\lambda_1 \lambda_2}, \quad
\lambda_3 + \lambda_4 - \left| \lambda_5 \right|
> - \sqrt{\lambda_1 \lambda_2}.
\ee
Inequalities~\eqref{huihu} and~\eqref{mnjoh}
are necessary and sufficient conditions for
boundedness-from-below when $\lambda_6 = \lambda_7 = 0$~\cite{deshpande};
they are necessary conditions when $\lambda_6$ and $\lambda_7$ are nonzero.

We may now return to inequality~\eqref{bjghiu},
which implies,
in principle,
many more necessary conditions for boundedness-from-below.
Setting for instance $\sin{\theta} = \cos{\theta}$ one concludes that
\be
2 h \left| \Re{\left[ \left( \lambda_6 + \lambda_7 \right)
    e^{i \alpha} \right]} \right|
< \frac{\lambda_1 + \lambda_2}{2} + \lambda_3 + \lambda_4 h^2
+ \Re{\left( \lambda_5 e^{2 i \alpha} \right)} h^2,
\ee
which must hold for any $h$ and $\alpha$.
Therefore~\cite{barroso},
\be
\label{mnjoh2}
2 \left| \lambda_6 + \lambda_7 \right|
< \frac{\lambda_1 + \lambda_2}{2} + \lambda_3 + \lambda_4
+ \left| \lambda_5 \right|.
\ee

We have numerically analyzed the BFB conditions by giving random
values to $\lambda_1$,
$\lambda_2$,
$\lambda_3$,
$\lambda_4$,
$\left| \lambda_5 \right|$,
$\left| \lambda_6 \right|$,
$\left| \lambda_7 \right|$,
$\arg{\left( \lambda_5^\ast \lambda_6 \lambda_7 \right)}$,
and $\arg{\left( \lambda_6^\ast \lambda_7 \right)}$ and then checking
whether the BFB conditions are met.
We have confirmed that the conditions~\eqref{huihu},
\eqref{mnjoh},
and~\eqref{mnjoh2} always hold.\footnote{The BFB conditions worked out
  in this subsection are,
  clearly,
  the ones valid at tree level.
  At loop level the BFB conditions change,
  see ref.~\cite{staub2}.}

\subsection{Procedure}

We consider the most general 2HDM
and purport to find out its ranges for $g_3$ and $g_4$.
We use the Higgs basis for the scalar doublets;
in that basis only $\phi_1^0$ has VEV
and therefore $\phi_1$ has the expression~\eqref{nuiho},
while
\be
\label{gjifo}
\phi_2 = \left( \begin{array}{c} C^+ \\ \left( \sigma_1 + i \sigma_2 \right)
  \left/ \sqrt{2} \right. \end{array} \right).
\ee
In equation~\eqref{gjifo},
$\sigma_1$ and $\sigma_2$ are real fields
and $C^+$ is the physical charged scalar of the 2HDM.
We emphasize that using the Higgs basis represents no lack of generality,
because both the unitarity and the BFB conditions are the same in any basis.

Since only $\phi_1$ has VEV,
the vacuum stability conditions are
$\mu_1 = - \lambda_1 v^2$ and $\mu_3 = - \lambda_6 v^2$~\cite{silvaCMU}.
The coupling $\mu_2$ in equation~\eqref{gmkho}
is unrelated to the parameters of $V_4$;
one may trade it for the charged-Higgs squared mass
$M_C = \mu_2 + \lambda_3 v^2$.
The mass terms of $H$,
$\sigma_1$,
and $\sigma_2$ are given by line~\eqref{jhuigo},
with~\cite{silvaCMU}
\be
\label{jghjoe}
M = \left( \begin{array}{ccc}
  2 \lambda_1 v^2 & 2 v^2\, \Re{\lambda_6} & - 2 v^2\, \Im{\lambda_6} \\
  2 v^2\, \Re{\lambda_6} &
  M_C + \left( \lambda_4 + \Re{\lambda_5} \right) v^2 &
  - v^2\, \Im{\lambda_5} \\
  - 2 v^2\, \Im{\lambda_6} &
  - v^2\, \Im{\lambda_5} &
  M_C + \left( \lambda_4 - \Re{\lambda_5} \right) v^2
\end{array} \right).
\ee
The matrix $M$ is diagonalized through equations~\eqref{MMM}--\eqref{RT}.

The three invariants  of $M$ are
\bs
\ba
I_1 (M) &=& 2 M_C + 2 \left( \lambda_1 + \lambda_4 \right) v^2, \\
I_2 (M) &=& M_C^2 + 2 \left( 2 \lambda_1 + \lambda_4 \right) v^2 M_C
+ \left( 4 \lambda_1 \lambda_4 + \lambda_4^2 - \left| \lambda_5 \right|^2
- 4 \left| \lambda_6 \right|^2 \right) v^4, \\
I_3 (M) &=& 2 \lambda_1 v^2 M_C^2 + 4 \left( \lambda_1 \lambda_4
- \left| \lambda_6 \right|^2 \right) v^4 M_C
\no & &
+ 2 \left[ \lambda_1 \lambda_4^2
  - \lambda_1 \left| \lambda_5 \right|^2
  - 2 \lambda_4 \left| \lambda_6 \right|^2
  + 2\, \Re{\left( \lambda_5^\ast \lambda_6^2 \right)} \right] v^6.
\ea
\es
We input parameters $\lambda_{1,2,\cdots,7}$ that
satisfy both the unitarity conditions and the BFB conditions
of subsections~\ref{uhi} and~\ref{uhi2},
respectively.\footnote{This method,
    where $\lambda_{1,2,\cdots,7}$ are used as input,
    tends to produce few points with either very low or very high
    scalar masses. Therefore we have supplemented it by another search
    in which we have directly used as input $M_{1,2,3,C}$.}
We also use the values of $M_1$ and $v$ in equations~\eqref{data}.
The two equations
\bs
\label{jhkpgfo}
\ba
M_1^3 - M_1^2 I_1(M) + M_1 I_2 (M) - I_3 (M) &=& 0, \\
\left[ M_1^2 I_1 (M) - 2 M_1 I_2 (M) + 3 I_3 (M) \right] \cos^2{\vartheta_1} & &
\no
+ M_{11} \left[ M_1 I_1 (M) - M_1^2 \right]
- \left( M^2 \right)_{11} M_1
- I_3 (M) &=& 0
\ea
\es
are quadratic in $M_C$.
By affirming the fact that both quadratic equations~\eqref{jhkpgfo}
must hold for the same value of $M_C$,
one is able to compute both $M_C$ and $\cos^2{\vartheta_1}$.
We thus get to know the full matrix $M$,
hence its eigenvalues $M_2$ and $M_3$
and its diagonalizing matrix $R$.

We require $\cos{\vartheta_1} > 0.9$.
We also compute the oblique parameter
\bs
\label{T_2HDM}
\ba
T &=& \frac{1}{16 \pi s_w^2 m_W^2} \left[
  s_1^2 F \left( M_C, M_1 \right)
  + \left( 1 - s_1^2 c_2^2 \right) F \left( M_C, M_2 \right)
  + \left( 1 - s_1^2 s_2^2 \right) F \left( M_C, M_3 \right)
  \right. \hspace*{7mm} \\ & & \left.
  - c_1^2 F \left( M_2, M_3 \right)
  - s_1^2 c_2^2 F \left( M_1, M_3 \right)
  - s_1^2 s_2^2 F \left( M_1, M_2 \right)
  \right] + T_\mathrm{singlets},
\ea
\es
where $T_\mathrm{singlets}$ is given by equation~\eqref{t}.
We require $-0.04 < T < 0.20$.

We have applied the method devised in ref.~\cite{silva-ivanov}
to guarantee that our assumed vacuum state
is indeed the state with the lowest possible value of the potential.
The method may be described as follows.
Let the matrix $\Lambda_E$ in equation~\eqref{lambdae}
have four eigenvalues $\Lambda_{0,1,2,3}$.
We already know from the BFB conditions that those eigenvalues
must be real and different from each other;
let us order them as $\Lambda_0 > \Lambda_1 > \Lambda_2 > \Lambda_3$.
Let the charged-Higgs squared mass be $M_C$;
define $\zeta \equiv 2 M_C \left/ v^2 \right.$.
Then, the assumed vacuum state is the global minimum of the potential
if either $\zeta > \Lambda_0$,
or $\Lambda_0 > \zeta > \Lambda_1$,
or $\Lambda_2 > \zeta > \Lambda_3$.
This test led us to discard about 10\% of our previous set of points.

The four-Higgs vertex is given by
\bs
\label{g4_2HDM}
\ba
g_4 &=&
\frac{\lambda_1 c_1^4}{8}
+ \frac{\lambda_2 s_1^4}{8}
+ \frac{\left( \lambda_3 + \lambda_4 \right) c_1^2 s_1^2}{4}
+ \frac{s_1^2 c_1^2 \left( c_3^2 - s_3^2 \right) \Re{\lambda_5}}{4}
- \frac{s_1^2 c_1^2 c_3 s_3\, \Im{\lambda_5}}{2}
\\ & &
+ \frac{s_1 c_1^3 \left( c_3\, \Re{\lambda_6} - s_3\, \Im{\lambda_6} \right)}{2}
+ \frac{s_1^3 c_1 \left( c_3\, \Re{\lambda_7} - s_3\, \Im{\lambda_7} \right)}{2}.
\ea
\es
The three-Higgs vertex is given by
\bs
\ba
g_3 &=& \frac{v}{\sqrt{2}} \left[
  \lambda_1 c_1^3
  + \left( \lambda_3 + \lambda_4 \right) s_1^2 c_1
  + s_1^2 c_1 \left( c_3^2 - s_3^2 \right) \Re{\lambda_5}
  - 2 s_1^2 c_1 c_3 s_3\, \Im{\lambda_5}
  \right. \\ & & \left.
  + 3 s_1 c_1^2 \left( c_3\, \Re{\lambda_6} - s_3\, \Im{\lambda_6} \right)
  + s_1^3 \left( c_3\, \Re{\lambda_7} - s_3\, \Im{\lambda_7} \right)
  \right].
\ea
\es

We also want to consider the $h_1 C^+ C^-$ vertex,
which may be relevant in the discovery of the charged scalar.
That vertex is given by
\be
V_4 = \cdots + h_1 C^+ C^- g_{1CC},
\ee
where,
in the 2HDM,
\be
\label{jhkhlp}
g_{1CC} = \sqrt{2} v \left(
c_1 \lambda_3 + s_1 c_3\, \Re{\lambda_7} - s_1 s_3\, \Im{\lambda_7} \right).
\ee

\subsection{Results}

\begin{figure}
\begin{center}
\epsfig{file=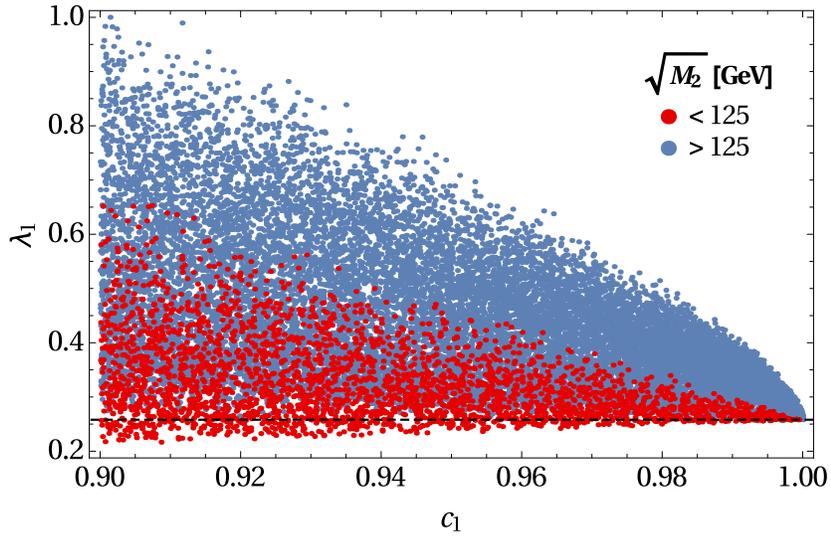,width=0.7\textwidth}
\end{center}
\caption{Scatter plots of $\lambda_1$
  \textit{versus} $\cos{\vartheta_1}$ in the 2HDM.
  The dashed line marks the value of $\lambda_1$ in the SM.
  The red points have $M_2 < M_1$.
  \label{fig20}}
\end{figure}

As we know from subsections~\ref{uhi} and~\ref{uhi2},
in general $\lambda_1$ can take any value in between $0$ and $4 \pi / 3$.
Once the constraint $\cos{\vartheta_1} > 0.9$ is imposed,
however,
$\lambda_1$ can be no larger than $\sim 1$;
this is illustrated in figure~\ref{fig20}.
The closer $\cos{\vartheta_1}$ is to 1,
the closer $\lambda_1$ must be to its SM value
$M_1 \left/ \left( 2 v^2 \right) \right. = 0.258$;
note that $\lambda_1$ is almost always larger than its SM value
when $\cos{\vartheta_1} > 0.9$;
the minimum value that we have obtained for $\lambda_1$
is 0.2135.

\begin{figure}
\begin{center}
\epsfig{file=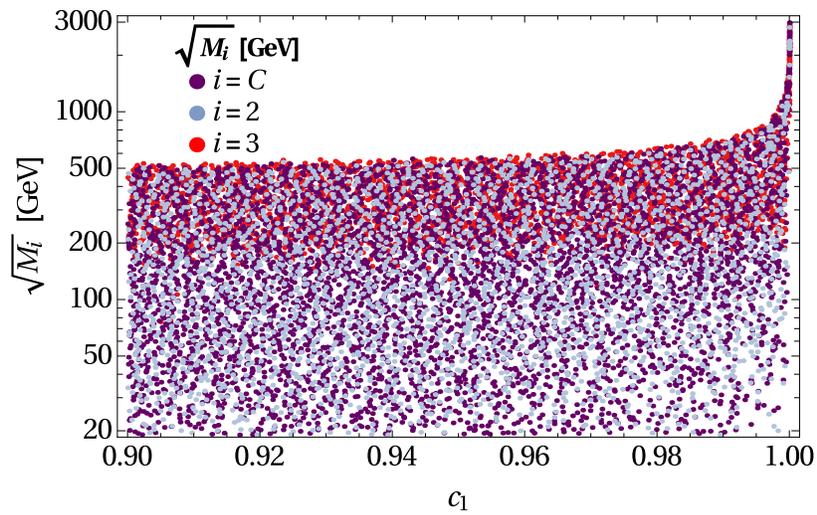,width=0.7\textwidth}
\end{center}
\caption{Scatter plots of the masses of the extra scalars of the 2HDM
  \textit{versus} $\cos{\vartheta_1}$.
  \label{fig21}}
\end{figure}

If $\cos{\vartheta_1} \lesssim 0.99$,
then the masses of the new scalar particles of the 2HDM,
namely $\sqrt{M_C}$,
$\sqrt{M_2}$,
and $\sqrt{M_3}$ can be no larger than $\sim 700$\,GeV;
if $\cos{\vartheta_1} \lesssim 0.95$,
they can be no larger than $\sim 550$\,GeV.
When $\cos{\vartheta_1}$ becomes close to 1,
the masses of the new scalar particles may reach O(TeV);
this is illustrated in figure~\ref{fig21}.

\begin{figure}
\begin{center}
\epsfig{file=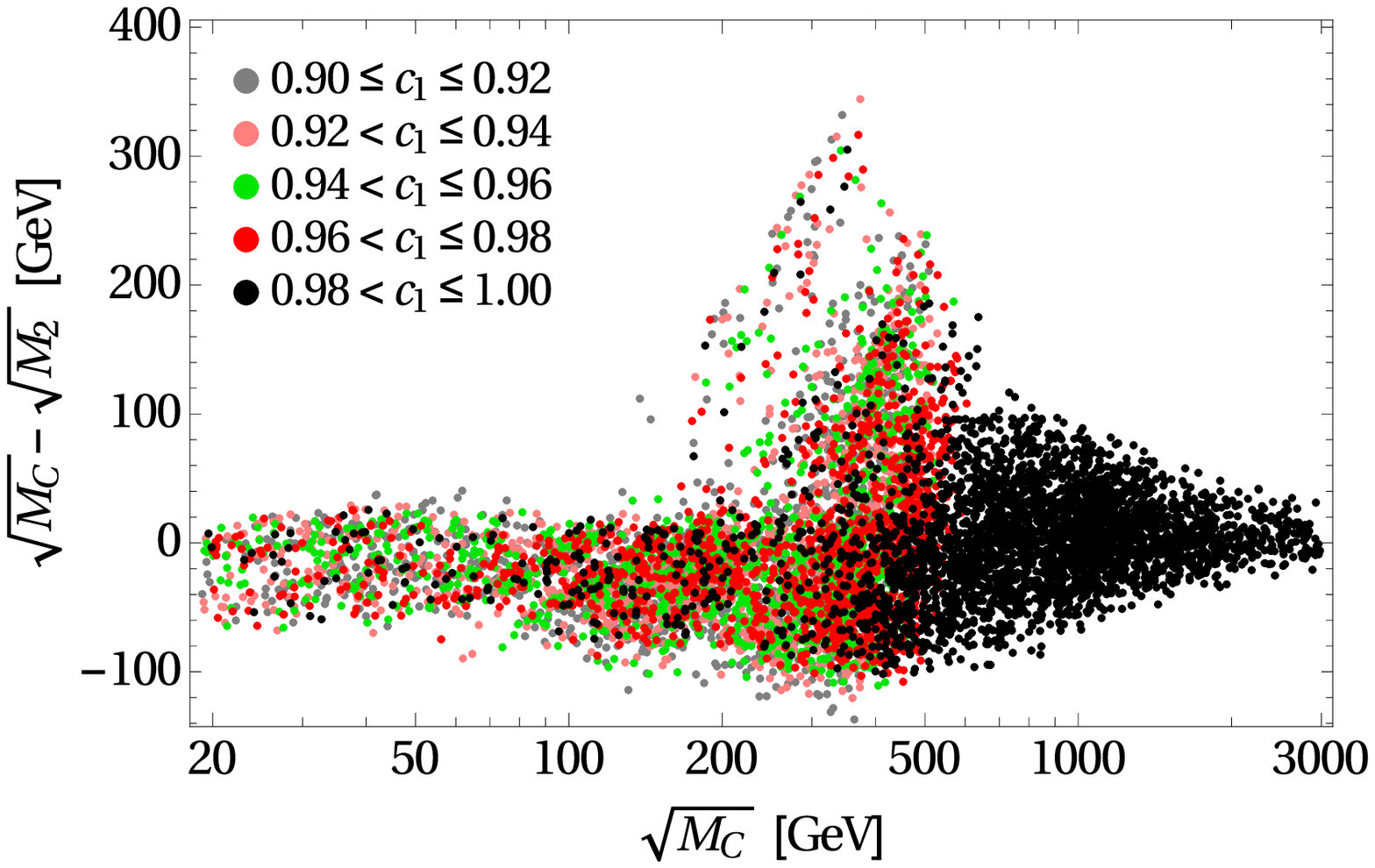,width=0.7\textwidth}
\end{center}
\caption{The difference between the mass of the charged scalar
  and the mass of the lightest non-SM neutral scalar
  \textit{versus} the mass of the charged scalar in the 2HDM.
  \label{fig22}}
\end{figure}

  One sees in figure~\ref{fig22}
  that $\sqrt{M_C}$ and $\sqrt{M_2}$ differ by at most $\sim$100\,GeV
  unless $200\,\mathrm{GeV} < \sqrt{M_C} < 500\,\mathrm{GeV}$.
  (Remember that by convention $M_2$ is always smaller than $M_3$,
  but they may be smaller than $M_1$.)

\begin{figure}
\begin{center}
\epsfig{file=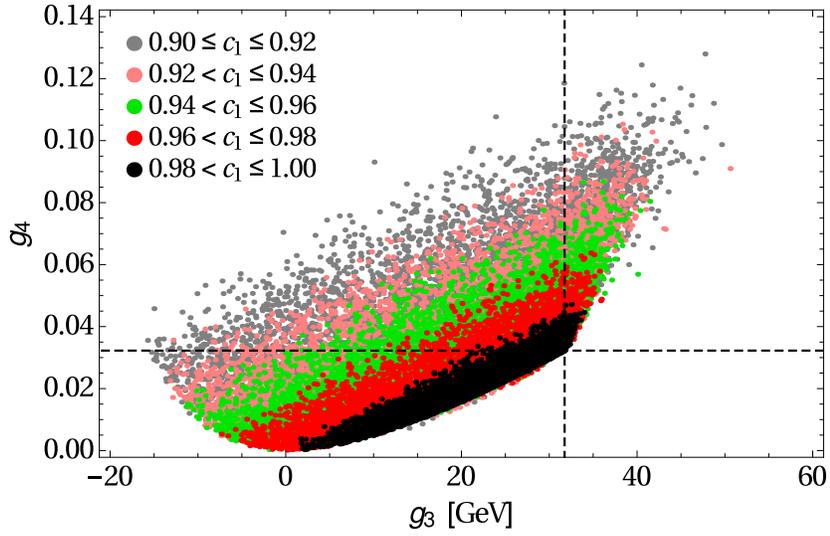,width=0.7\textwidth}
\end{center}
\caption{Scatter plot of the four-Higgs coupling $g_4$ \textit{versus}\/
  the three-Higgs coupling $g_3$ in the 2HDM,
  for various values of $c_1$.
  The dashed lines mark the values of both couplings in the SM.
  \label{fig24}}
\end{figure}

We now come to the predictions for $g_3$ and $g_4$ in the 2HDM,
which are depicted in figure~\ref{fig24}.
One sees that
  $g_3$
  in the 2HDM has a range only slightly larger than in the SM2S,
  while $g_4$ in the 2HDM is much more restricted than in the SM2S;
  $g_4 \left/ g_4^\mathrm{SM} \right. \lesssim 4$ in the 2HDM
  but $g_4 \left/ g_4^\mathrm{SM} \right. \lesssim 15$ in the SM2S.
An interesting feature is that $g_3$ may be zero or even negative,
\textit{i.e.}\ it may have sign opposite to the one in the SM.
(We recall that the sign of $g_3$ is measured relative to the sign of $c_1$;
we arrange that $c_1$ is always positive.)
On the other hand,
$g_4$ is always positive because of the boundedness from below of the potential.

\begin{figure}
\begin{center}
\epsfig{file=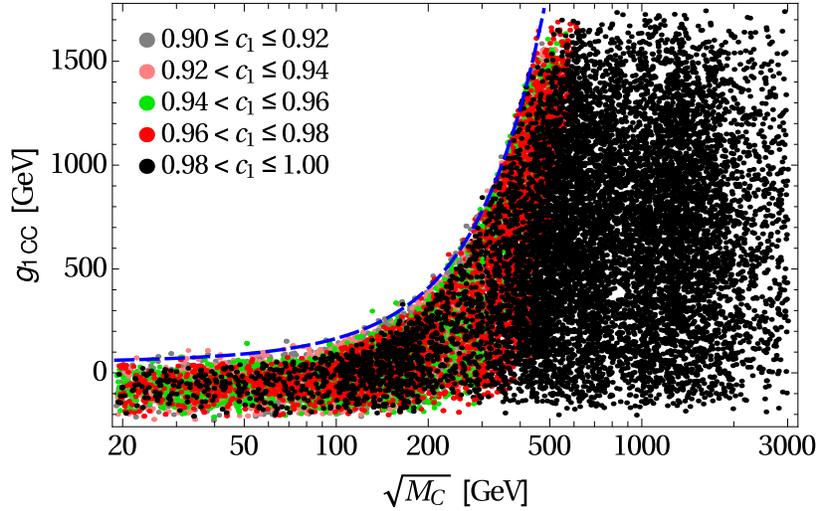,width=0.7\textwidth}
\end{center}
\caption{Scatter plot of the $h_1 C^+ C^-$ coupling $g_{1CC}$
  \textit{versus}\/ the mass of the charged scalars $C^\pm$
  in the 2HDM.
    The blue line with equation
    $g_{1CC} \! \left/ \mathrm{GeV} \right.
    = 48.5 + 0.54 \left( \sqrt{M_C} \! \left/ \mathrm{GeV} \right. \right)
    + 0.0063 \left( M_C \! \left/ \mathrm{GeV}^2 \right. \right)$
    marks the approximate boundary of the allowed region.
  \label{fig25}}
\end{figure}

In figure~\ref{fig25} we depict the coupling $g_{1CC}$
of the 125\,GeV neutral scalar to a pair of charged scalars in the 2HDM.
  One sees that that coupling is in between -200\,GeV
  and 1,700\,GeV.
  The expression for $g_{1CC}$ in equation~\eqref{jhkhlp}
  is strongly dominated by the first term in the right-hand side
  because $c_1 \gg s_1$.
  The preference for positive values of $g_{1CC}$ observed in figure~\ref{fig25}
  occurs because $-2 \lesssim \lambda_3 \lesssim 7$
  in the 2HDM with the constraint $c_1 > 0.9$ enforced.

\section{The two-Higgs-doublet model plus one singlet} \label{sec:2HDMsinglet}

We consider in this section the two-Higgs-doublet model
with the addition of one real $SU(2) \times U(1)$-invariant
scalar field $S$.
We assume a symmetry $S \to -S$.
As a shorthand,
we shall dub this model the 2HDM1S
(other authors use just 2HDMS~\cite{arhrib}).
The quartic part of the scalar potential is
\bs
\label{buioh}
\ba
V_4 &=&
\frac{\lambda_1}{2} \left( \phi_1^\dagger \phi_1 \right)^2
+ \frac{\lambda_2}{2} \left( \phi_2^\dagger \phi_2 \right)^2
+ \lambda_3\, \phi_1^\dagger \phi_1\, \phi_2^\dagger \phi_2
+ \lambda_4\, \phi_1^\dagger \phi_2\, \phi_2^\dagger \phi_1
\hspace*{7mm} \\ & &
+ \left[
  \frac{\lambda_5}{2} \left( \phi_1^\dagger \phi_2 \right)^2
  + \lambda_6\, \phi_1^\dagger \phi_1\, \phi_1^\dagger \phi_2
  + \lambda_7\, \phi_2^\dagger \phi_2\, \phi_1^\dagger \phi_2
  + \mathrm{H.c.}
  \right]
\\ & &
+ \frac{\psi}{2}\, S^4
\\ & &
+ S^2 \left(
\xi_1\, \phi_1^\dagger \phi_1 + \xi_2\, \phi_2^\dagger \phi_2
+ \xi_3\, \phi_1^\dagger \phi_2 + \xi_3^\ast\, \phi_2^\dagger \phi_1 \right).
\ea
\es

\subsection{Bounded-from-below conditions} \label{sec:fghdj}

Deriving necessary and sufficient BFB conditions
for even a rather simple potential like the one in equation~\eqref{buioh}
is a notoriously difficult problem~\cite{igormargarete}.
If $V_4$ were negative for some possible values of $S^2$,
$\phi_1^\dagger \phi_1$,
$\phi_2^\dagger \phi_2$,
and $\phi_1^\dagger \phi_2$,
then $V_4$ would tend to $- \infty$ upon multiplication of those four values
by an ever-larger positive constant.
Therefore,
we want $V_4$ to be positive for all possible values of $S^2$,
$\phi_1^\dagger \phi_1$,
$\phi_2^\dagger \phi_2$,
and $\phi_1^\dagger \phi_2$.
In order to guarantee this,
we proceed in the following fashion.

\paragraph{Necessary condition 1:}
When $S^2 = 0$,
equation~\eqref{buioh} reduces to its first two lines,
\textit{i.e.}\ to the quartic potential of the 2HDM.
Therefore,
one must require the fulfilment of the conditions of subsection~\ref{uhi2},
\textit{viz.}\ the four conditions
in between equations~\eqref{lambdae} and~\eqref{gohpu}.

\paragraph{Necessary condition 2:}
When $\phi_1^\dagger \phi_2 = 0$,
\be
V_4 = \frac{1}{2}
\left( \begin{array}{ccc}
  \phi_1^\dagger \phi_1 & \phi_2^\dagger \phi_2 & S^2
\end{array} \right)
\left( \begin{array}{ccc}
  \lambda_1 & \lambda_3 & \xi_1 \\
  \lambda_3 & \lambda_2 & \xi_2 \\
  \xi_1 & \xi_2 & \psi
\end{array} \right)
\left( \begin{array}{c}
  \phi_1^\dagger \phi_1 \\ \phi_2^\dagger \phi_2 \\ S^2
\end{array} \right).
\ee
Since $\phi_1^\dagger \phi_1$,
$\phi_2^\dagger \phi_2$,
and $S^2$ are positive definite quantites,
we must require~\cite{kannike,kannike2}
\bs
\ba
\psi &>& 0, \label{psipsipsi} \\
\lambda_1 &>& 0, \\
\lambda_2 &>& 0, \\
A_1 \equiv \xi_1 + \sqrt{\lambda_1 \psi} &>& 0, \\
A_2 \equiv \xi_2 + \sqrt{\lambda_2 \psi} &>& 0, \\
A_3 \equiv \lambda_3 + \sqrt{\lambda_1 \lambda_2} &>& 0, \\
\sqrt{\lambda_1 \lambda_2 \psi} + \xi_2 \sqrt{\lambda_1}
+ \xi_1 \sqrt{\lambda_2} + \lambda_3 \sqrt{\psi}
+ \sqrt{2 A_1 A_2 A_3} &>& 0.
\ea
\es

  After enforcing the necessary condition~1,
  we know that $V_4 > 0$
  when only the first two lines of the potential~\eqref{buioh} exist;
  after enforcing the inequality~\eqref{psipsipsi},
  we know that $V_4 > 0$ when only the third line exists.
If we guarantee that the fourth line
of the potential~\eqref{buioh} is always positive too,
then we will be sure that $V_4$ is always positive.
We therefore have the following\footnote{We thank Igor Ivanov
  for pointing out this sufficient condition to us.}

\paragraph{Sufficient condition:}
If,
besides the two necessary conditions,
\bs
\label{gjhio}
\ba
\xi_1 + \xi_2 &>& 0, \\
\xi_1 \xi_2 - \left| \xi_3 \right|^2 &>& 0,
\ea
\es
then $V_4$ is BFB.

Among the sets of parameters of the potential~\eqref{buioh}
that we have randomly generated,
there were some that met both the two necessary conditions
and the sufficient conditions~\eqref{gjhio};
we have used those sets of parameters.
There were many other sets that satisfied the two necessary conditions
but did not meet the sufficient conditions~\eqref{gjhio};
for those sets,
we have numerically found the absolute minimum of $V_4$.
We have done this by using $S^2 = 1$ together with equations~\eqref{r2}
and by minimizing $V_4$ in the domain $r^2 > 0$,
$0 \le \theta \le \pi/2$,
$0 \le h \le 1$,
and $0 \le \alpha < 2 \pi$.
If the minimum of $V_4$ is positive,
then the set of input parameters is good,
else the set of input parameters is bad and one must discard it.

\subsection{Unitarity conditions}

There are the same five scattering channels as in the 2HDM,
\textit{cf.}\ subsection~\ref{uhi};
but the channel $Q = T_3 = 0$ has an additional scattering state $S^2$.
Additionally,
there are two extra scattering channels:
\begin{itemize}
\item The channel $Q = 1,\ T_3 = 1/2$ with the two states $aS$ and $cS$.
\item The channel $Q = 0,\ T_3 = -1/2$ with the two states $bS$ and $dS$.
\end{itemize}
Both these channels produce a scattering matrix
\be
\mathcal{M}_4 = 2 \left( \begin{array}{cc}
  \xi_1 & \xi_3 \\ \xi_3^\ast & \xi_2
\end{array} \right).
\ee
Channels~\ref{channel10} and~\ref{channel20} of subsection~\ref{uhi}
again produce the scattering matrix~\eqref{buigpj}.
Channel~\eqref{channel30} produces that matrix
together with the additional eigenvalue $I$ of equation~\eqref{I}.
Channel~\eqref{channel20} produces the scattering matrix~\eqref{jhui}.
Finally,
channel~\ref{channel50} has the additional scattering state $S^2$
and therefore,
instead of producing both the matrix $\mathcal{M}_1$ of equation~\eqref{jhui}
and the matrix $\mathcal{M}_2$ of equation~\eqref{uhiooo},
it produces $\mathcal{M}_1$ together with
\be
\label{sjhkg}
\mathcal{M}_2^\prime = \left( \begin{array}{cc}
  6 \psi & \sqrt{2}\, \bar \xi^T \\
  \sqrt{2}\, \bar \xi & \mathcal{M}_2
\end{array} \right),
\quad \mbox{where} \quad
\bar \xi = \left( \begin{array}{c}
  \xi_1 + \xi_2 \\ 2\, \Re{\xi_3} \\ - 2\, \Im{\xi_3} \\
  \xi_1 - \xi_2
\end{array} \right).
\ee
Thus,
the unitarity conditions for the 2HDM1S are the following:
both $\left| I \right|$ and the moduli of all the eigenvalues
of the $2 \times 2$ matrix~$\mathcal{M}_4$,
of the $3 \times 3$ matrix~$\mathcal{M}_3$,
of the $4 \times 4$ matrix~$\mathcal{M}_1$,
and of the $5 \times 5$ matrix~$\mathcal{M}_2^\prime$
must be smaller than $4 \pi$.

\subsection{Procedure} \label{proc2HDM1S}

Just as in the previous section,
we utilize the Higgs basis for the two doublets,
\textit{i.e.}\ equations~\eqref{nuiho} and~\eqref{gjifo}.
We also write $S = w + \sigma$,
where $w$ is the VEV of the scalar $S$ and $\sigma$ is a field.
The mass terms of the scalars are
\be
V = \cdots + M_C C^- C^+ + \frac{1}{2}
\left( \begin{array}{cccc} H & \sigma_1 & \sigma_2 & \sigma \end{array} \right)
M
\left( \begin{array}{c} H \\ \sigma_1 \\ \sigma_2 \\ \sigma \end{array} \right),
\ee
with
\be
M = \left( \begin{array}{cccc}
  2 \lambda_1 v^2 &
  2 v^2\, \Re{\lambda_6} &
  - 2 v^2\, \Im{\lambda_6} &
  2 \sqrt{2} v w \xi_1 \\
  2 v^2\, \Re{\lambda_6} &
  M_C + \left( \lambda_4 + \Re{\lambda_5} \right) v^2 &
  - v^2\, \Im{\lambda_5} &
  2 \sqrt{2} v w\, \Re{\xi_3} \\
  - 2 v^2\, \Im{\lambda_6} &
  - v^2\, \Im{\lambda_5} &
  M_C + \left( \lambda_4 - \Re{\lambda_5} \right) v^2 &
  - 2 \sqrt{2} v w\, \Im{\xi_3} \\
  2 \sqrt{2} v w \xi_1 &
  2 \sqrt{2} v w\, \Re{\xi_3} &
  - 2 \sqrt{2} v w\, \Im{\xi_3} &
  4 \psi w^2
\end{array} \right),
\label{vjgugo}
\ee
\textit{cf.}\ equation~\eqref{jghjoe}.
One diagonalizes $M$ as
\bs
\ba
M &=& R^T\, \mathrm{diag} \left( M_1,\, M_2,\, M_3,\, M_4 \right) R,
\\*[1mm]
\left( \begin{array}{c} H \\ \sigma_1 \\ \sigma_2 \\ \sigma \end{array} \right)
&=& R^T
\left( \begin{array}{c} h_1 \\ h_2 \\ h_3 \\ h_4 \end{array} \right),
\ea
\es
where $R$ is a $4 \times 4$ orthogonal matrix.
The squared mass $M_1$ is given by equation~\eqref{M1}.
Without loss of generality,
$M_2 < M_3 < M_4$.
Just as in the previous sections,
we require
\be
\label{bmkgp}
R_{11} \equiv c_1 > 0.9.
\ee
The expression for the oblique parameter $T$ is~\cite{vikings}
\bs
\ba
T &=& \frac{1}{16 \pi s_w^2 m_W^2} \left\{
\sum_{k=1}^4 \left[ \left( R_{k2} \right)^2 + \left( R_{k3} \right)^2 \right]
F \left( M_C,\, M_k \right)
\right. \\ & &
- \sum_{k=1}^3\, \sum_{k^\prime = k + 1}^4 \left( R_{k2} R_{k^\prime 3}
- R_{k^\prime 2} R_{k3} \right)^2 F \left( M_k,\, M_{k^\prime} \right)
\\ & & + 3 \sum_{k=2}^4 \left( R_{k1} \right)^2 \left[
  F \left( M_k,\, m_Z^2 \right) - F \left( M_k,\, m_W^2 \right) \right]
\\ & & \left.
+ 3 \left( c_1^2 - 1 \right) \left[
  F \left( M_1,\, m_Z^2 \right) - F \left( M_1,\, m_W^2 \right) \right]
\right\},
\ea
\es
and we demand $-0.04 < T < 0.20$.

We input random values for the 15 real parameters $M_C$,
$\lambda_{1,2,3,4}$,
$\left| \lambda_{5,6,7} \right|$,
$\psi$,
$\xi_{1,2}$,
$\left| \xi_3 \right|$,
$\arg{\left( \lambda_5^\ast \lambda_6 \lambda_7 \right)}$,
$\arg{\left( \lambda_6^\ast \lambda_7 \right)}$,
and $\arg{\left( \lambda_6^\ast \xi_3 \right)}$.
We moreover input $M_1$ and $v^2$ given in equations~\eqref{data}.
Then,
\begin{enumerate}
\item We require the input parameters to satisfy
  the BFB conditions of subsection~\ref{sec:fghdj}---this may imply
  a numerical minimization of $V_4$ to check that $V_4 > 0$.
\item We require the input parameters to satisfy
  the unitarity conditions written after equation~\eqref{sjhkg}.
\item We compute the VEV $w$ from the condition that $M_1$
  should be an eigenvalue of the matrix $M$.
\item We enforce the conditions in appendix~\ref{appendixC}.
    They guarantee that the vacuum state
    with $v = 174$\,GeV and $w \neq 0$ has a lower value of the potential
    than all the other possible stability points of the potential.
\item We compute the full matrix $M$,
  its eigenvalues $M_{2,3,4}$,
  and its diagonalizing matrix $R$;
  we choose the overall sign of $R$ such that $R_{11} \equiv c_1 > 0$.
\item We impose both the condition~\eqref{bmkgp}
  and the condition that the oblique parameter $T$
  is within its experimental bounds.
\item We compute the couplings
  \bs
  \label{g3g3g3}
  \ba
  g_3 &=& \frac{v}{\sqrt{2}} \left\{ \lambda_1 c_1^3
  + \left( \lambda_3 + \lambda_4 \right) c_1
  \left[ \left( R_{12} \right)^2 + \left( R_{13} \right)^2 \right] \right.
  \\ & &
  + c_1
  \left[ \left( R_{12} \right)^2 - \left( R_{13} \right)^2 \right] \Re{\lambda_5}
  - 2 c_1 R_{12} R_{13}\, \Im{\lambda_5}
  \\ & & \left.
  + 3 c_1^2 \left( R_{12} \Re{\lambda_6} - R_{13} \Im{\lambda_6} \right)
  + \left[ \left( R_{12} \right)^2 + \left( R_{13} \right)^2 \right]
  \left( R_{12} \Re{\lambda_7} - R_{13} \Im{\lambda_7} \right) \right\}
  \\ & & + 2 \psi w \left( R_{14} \right)^3
  + \xi_1 c_1 R_{14} \left( w c_1 + \sqrt{2} v R_{14} \right)
  + \xi_2 w R_{14}
  \left[ \left( R_{12} \right)^2 + \left( R_{13} \right)^2 \right]
  \hspace*{7mm}
  \\ & &
  + \sqrt{2} R_{14} \left( v R_{14} + \sqrt{2} w c_1 \right)
  \left( R_{12} \Re{\xi_3} - R_{13} \Im{\xi_3} \right),
  \ea
  \es
  \bs
  \ba
  g_4 &=& \frac{\lambda_1 c_1^4}{8} + \frac{\lambda_2}{8}
  \left[ \left( R_{12} \right)^2 + \left( R_{13} \right)^2 \right]^2
  + \frac{\lambda_3 + \lambda_4}{4}\, c_1^2
  \left[ \left( R_{12} \right)^2 + \left( R_{13} \right)^2 \right]
  \\ & & + \frac{\Re{\lambda_5}}{4}\, c_1^2
  \left[ \left( R_{12} \right)^2 - \left( R_{13} \right)^2 \right]
  - \frac{\Im{\lambda_5}}{2}\, c_1^2 R_{12} R_{13}
  \\ & & + \frac{c_1^3}{2}
  \left( R_{12}\, \Re{\lambda_6} - R_{13}\, \Im{\lambda_6} \right) + \frac{c_1
    \left[ \left( R_{12} \right)^2 + \left( R_{13} \right)^2 \right]}{2}
  \left( R_{12}\, \Re{\lambda_7} - R_{13}\, \Im{\lambda_7} \right)
  \hspace*{7mm} \\ & &
  + \frac{\psi}{2} \left( R_{14} \right)^4
  \\ & &
  + \left( R_{14} \right)^2 \left\{
  \frac{\xi_1 c_1^2}{2}
  + \frac{\xi_2}{2}
  \left[ \left( R_{12} \right)^2 + \left( R_{13} \right)^2 \right]
  + c_1 \left( R_{12}\, \Re{\xi_3} - R_{13}\, \Im{\xi_3} \right) \right\},
  \ea
  \es
  \ba
  \label{hgytid}
  g_{1CC} &=& \sqrt{2} v \left( c_1 \lambda_3 + R_{12}\, \Re{\lambda_7}
  - R_{13}\, \Im{\lambda_7} \right) + 2 w \xi_2 R_{14}.
  \ea
\end{enumerate}

\subsection{Results}

\begin{figure}
\begin{center}
\epsfig{file=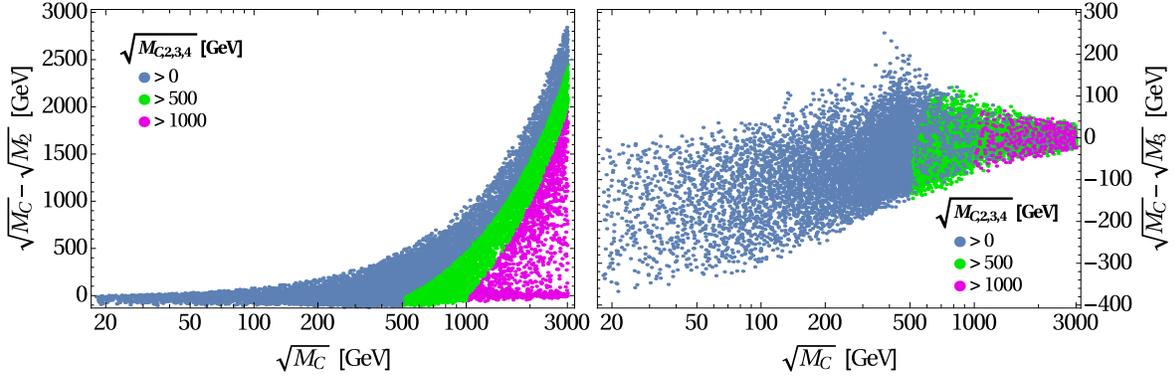,width=1.0\textwidth}
\end{center}
\caption{The differences between the masses
  of the two lightest non-SM neutral scalars and the mass of the charged scalar
  \textit{versus} the mass of the charged scalar in the 2HDM1S.
  Green points have all the scalars with mass larger than 500\,GeV;
  magenta points have all the scalars with mass larger than 1\,TeV.
  \label{fig30}}
\end{figure}

In figure~\ref{fig30} we have plotted the differences
among the masses of the scalars against the mass of the charged scalar.
One sees that $\sqrt{M_C}$ and $\sqrt{M_3}$
cannot be more than $\sim 300$\,GeV from each other,
but $\sqrt{M_2}$ may be
much
smaller than both of them.

\begin{figure}
\begin{center}
\epsfig{file=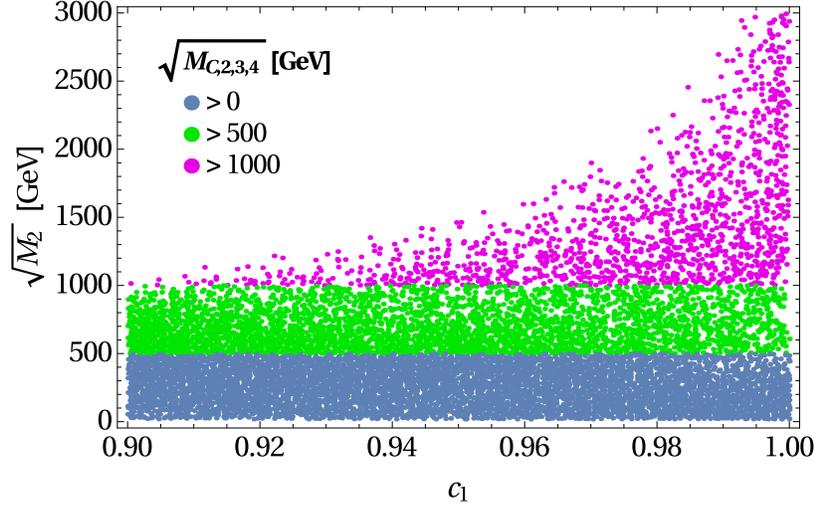,width=0.7\textwidth}
\end{center}
\caption{The mass of the lightest non-SM neutral scalar
  \textit{versus} $R_{11}$ in the 2HDM1S.
  Green points have all the scalars with mass larger than 500\,GeV;
  magenta points have all the scalars with mass larger than 1\,TeV.
  \label{fig31}}
\end{figure}

In figure~\ref{fig31} we present a scatter plot of the
mass of the lightest non-SM neutral scalar
against $c_1$.
One sees that,
contrary to what happens in the 2HDM (\textit{cf}.\ figure~\ref{fig21}),
$\sqrt{M_2}$ may reach 1\,TeV even when $c_1$ is as low as 0.9.

\begin{figure}
\begin{center}
\epsfig{file=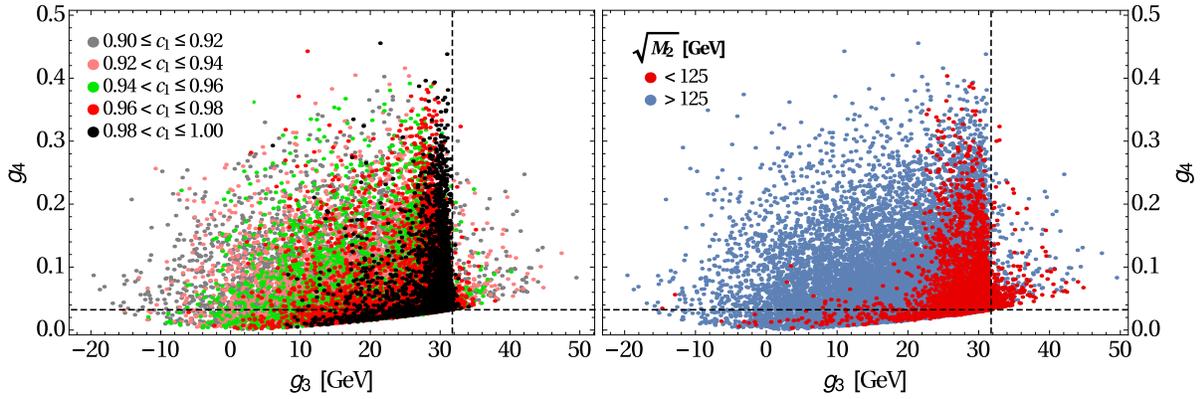,width=1.0\textwidth}
\end{center}
\caption{In the left panel,
  the four-Higgs coupling $g_4$
  \textit{versus} the three-Higgs coupling $g_3$ in the 2HDM1S
  for various values of $c_1$.
  The right panel contains the same points as the left panel
  but with different colours depending on whether $M_2$
  is larger or smaller than $M_1$.
  The dashed lines mark the values of the couplings in the SM.
  \label{fig32}}
\end{figure}

We depict in figure~\ref{fig32} the three- and four-Higgs couplings
$g_3$ and $g_4$ in the 2HDM1S.
  The main difference relative to the 2HDM
  (\textit{cf.}\ figure~\ref{fig24})
  is that $g_4$ may be much higher,
  just as in the SM2S.
In the 2HDM1S there is no clear correlation between $g_3$ and $g_4$.

\begin{figure}
\begin{center}
\epsfig{file=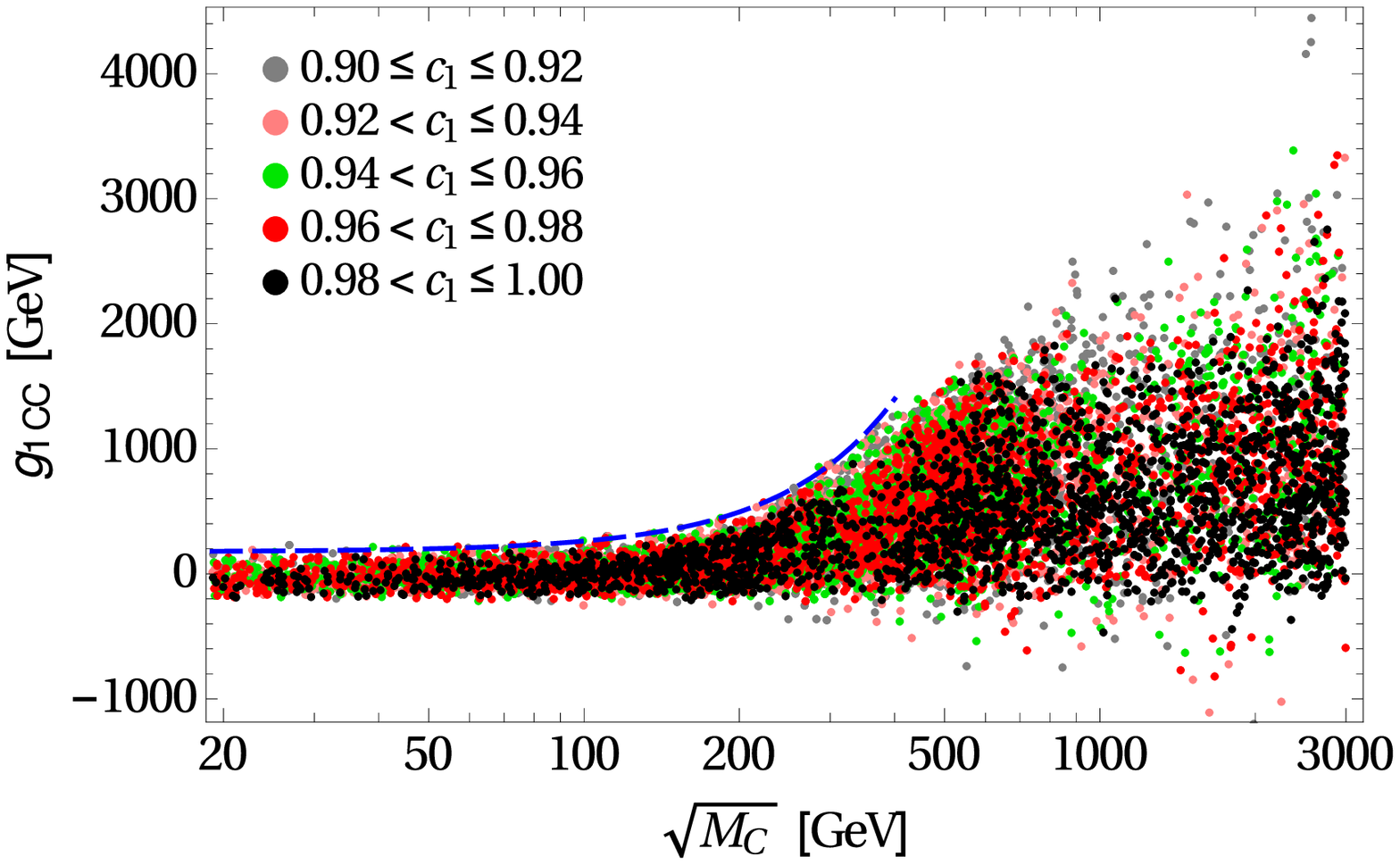,width=0.7\textwidth}
\end{center}
\caption{Scatter plot of $g_{1CC}$
  \textit{versus}\/ the mass of the charged scalars $C^\pm$ in the 2HDM1S.
    The blue line with equation
    $g_{1CC} \! \left/ \mathrm{GeV} \right.
    = 174.9 + 0.138 \left( \sqrt{M_C} \! \left/ \mathrm{GeV} \right. \right)
    + 0.0073 \left( M_C \! \left/ \mathrm{GeV}^2 \right. \right)$
    marks the approximate boundary of the allowed region
    when $\sqrt{M_C} < 400$\,GeV.
  \label{fig33}}
\end{figure}

In figure~\ref{fig33} we have plotted the $h_1 C^+ C^-$ coupling $g_{1CC}$.
That coupling in the 2HDM1S
  may be more than two times larger than in the 2HDM;
very large values of $g_{1CC}$ occur even for $c_1$ very close to 1.
This is because the right-hand side
  of equation~\eqref{hgytid}
  may be dominated by its fourth term when $w \gg v$.
  The first term displays the same behaviour
  as the corresponding term in the 2HDM,
  \textit{viz.}\ it is usually positive and no larger than 1,500\,GeV,
  but it is often overwhelmed by the fourth term.

\section{Conclusions}

In this paper we have emphasized that
both the bounded-from-below (BFB) conditions
and the unitarity conditions
for the two-Higgs-doublet model (2HDM)
are invariant under a change of the basis used for the two doublets.
Therefore,
one may implement those conditions directly in the Higgs basis,
\textit{viz.}\ the basis where only one doublet has vacuum expectation value.
This procedure allows one to extract bounds
on the masses and couplings of the scalar particles of the most general 2HDM,
disregarding any symmetry that a particular 2HDM may possess.
We have focussed
  on the three couplings $g_3 \left( h_1 \right)^3$,
  $g_4 \left( h_1 \right)^4$,
  and $g_{1CC} h_1 C^+ C^-$,
  where $h_1$ is the observed neutral scalar with mass 125\,GeV
  and $C^\pm$ are the charged scalars of the 2HDM.

\begin{figure}
\begin{center}
\epsfig{file=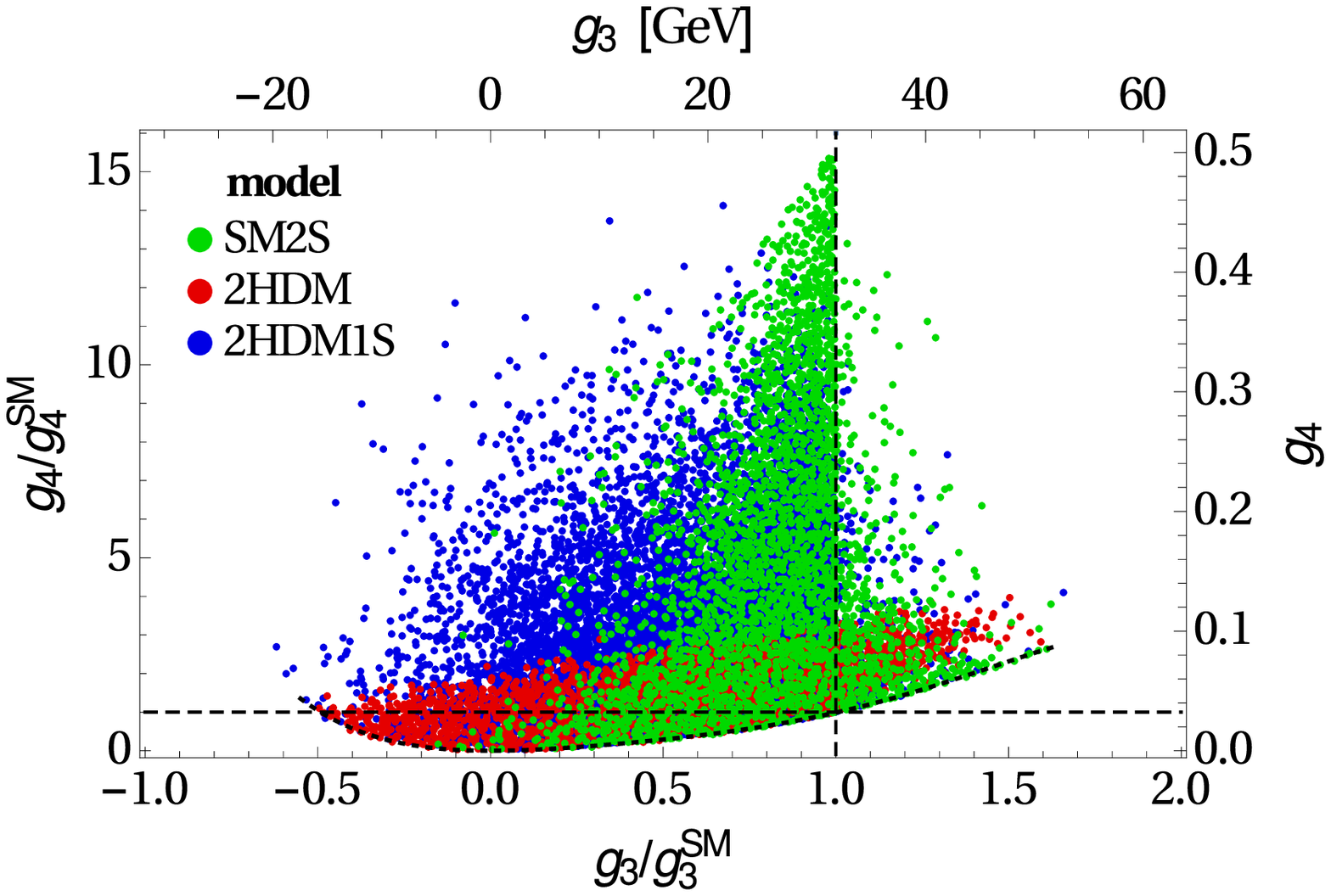,width=0.7\textwidth}
\end{center}
\caption{Scatter plot of $g_4 \left/ g_4^\mathrm{SM} \right.$
  \textit{versus} $g_3 \left/ g_3^\mathrm{SM} \right.$
  in the three models that we have studied.
  The dashed lines mark the SM values $g_3 \left/ g_3^\mathrm{SM} \right.
  = g_4 \left/ g_4^\mathrm{SM} \right. = 1$.
  The dotted line,
  with equation $g_4 \left/ g_4^\mathrm{SM} \right.
  = 2.06 \left( g_3 \left/ g_3^\mathrm{SM} \right. \right)^2
  - 2.84 \left( g_3 \left/ g_3^\mathrm{SM} \right. \right)^3
  + 2.44 \left( g_3 \left/ g_3^\mathrm{SM} \right. \right)^4
  - 0.67 \left( g_3 \left/ g_3^\mathrm{SM} \right. \right)^5$,
  marks the approximate boundary of the allowed region
  for $-0.6 < g_3 \left/ g_3^\mathrm{SM} \right. < 1.6$.
  \label{fig41}}
\end{figure}

  We have utilized the same procedure for two other models,
  namely the Standard Model with the addition of two real singlets (SM2S)
  and the two-Higgs-doublet model
  with the addition of one real singlet (2HDM1S),
  in both cases with reflection symmetries acting on each of the singlets.
  We have found,
  for instance,
  that:
  \begin{itemize}
  \item The coupling $g_3$ may,
    in both the 2HDM and the 2HDM1S,
    have sign opposite to the one in the SM.
    On the other hand,
    in any of the three models that we have studied,
    $\left| g_3 \right|$ can hardly be much larger than in the SM.
  \item The coupling $g_4$,
    which is always positive because of BFB,
    may for all practical purposes be equal to zero in all the three models.
    (As a matter of fact,
    $g_3 = g_4 = 0$ is possible in all three models.)
    But it may also be much larger than in the SM.
    A distinguished feature is that $g_4$ may be much larger
    (up to $g_4 \sim 0.5$)
    in the models containing singlets than in the 2HDM,
    wherein it can at best reach $g_4 \sim 0.13$.
  \item The coupling $g_{1CC}$ may be of order TeV,
    but only when the mass of $C^\pm$ exceeds 300\,GeV;
    in general,
    a positive $g_{1CC}$ may be larger for higher masses of $C^\pm$,
    but $g_{1CC}$ may also be negative for any $C^\pm$ mass.
    Moreover,
    $g_{1CC}$ may be more than two times larger
    (either positive or negative)
    in the 2HDM1S than in the 2HDM.
  \end{itemize}
A comparison of the predictions of the three models for $g_3$ and $g_4$
is depicted in figure~\ref{fig41}.

We emphasize that our method may be used to obtain bounds and/or correlations
among other parameters and/or observables of these models.
Unfortunately,
it may be difficult to generalize our work to more complicated models,
both because they may contain too many parameters
and because it is very difficult to derive full BFB conditions
for even rather simple models.

\vspace*{5mm}

\noindent \textbf{Acknowledgements}:
L.L.\ thanks Pedro Miguel Ferreira and Igor Ivanov,
and D.J.\ thanks Art\={u}ras Acus,
for useful discussions.
D.J.\ thanks the Lithuanian Academy of Sciences for support
through the project DaFi2018.
The work of L.L.\ is supported by the Portuguese
\textit{Funda\c c\~ao para a Ci\^encia e a Tecnologia}\/
through the projects CERN/FIS-NUC/0010/2015,
CERN/FIS-PAR/0004/2017,
and UID/FIS/00777/2013;
those projects are partly funded by POCTI (FEDER),
COMPETE,
QREN,
and the European Union.

\begin{appendix}

\setcounter{equation}{0}
\renewcommand{\theequation}{A\arabic{equation}}

\section{The Higgs Singlet Model}
\label{appendixA}

The Higgs Singlet Model (HSM) is the Standard Model with the addition
of one real scalar singlet $S$.
We furthermore assume a symmetry $S \to - S$.
The scalar potential
\be
V = \mu \phi_1^\dagger \phi_1 + m^2 S^2
+ \frac{\lambda}{2} \left( \phi_1^\dagger \phi_1 \right)^2
+ \frac{\psi}{2}\, S^4 + \xi S^2 \phi_1^\dagger \phi_1
\ee
has just five parameters $\mu$,
$m^2$,
$\lambda$,
$\psi$,
and $\xi$.
The bounded-from-below (BFB) conditions are
\be
\label{BFB}
\lambda > 0, \quad \psi > 0, \quad \xi > - \sqrt{\lambda \psi}.
\ee
The unitarity conditions are
\be
\label{unit}
\left| \lambda \right| < 4 \pi, \quad \left| \xi \right| < 2 \pi, \quad
\left| 3 \lambda + 6 \psi + \sqrt{\left( 3 \lambda - 6 \psi \right)^2
  + 16 \xi^2} \right| < 8 \pi.
\ee

\begin{figure}
\begin{center}
\epsfig{file=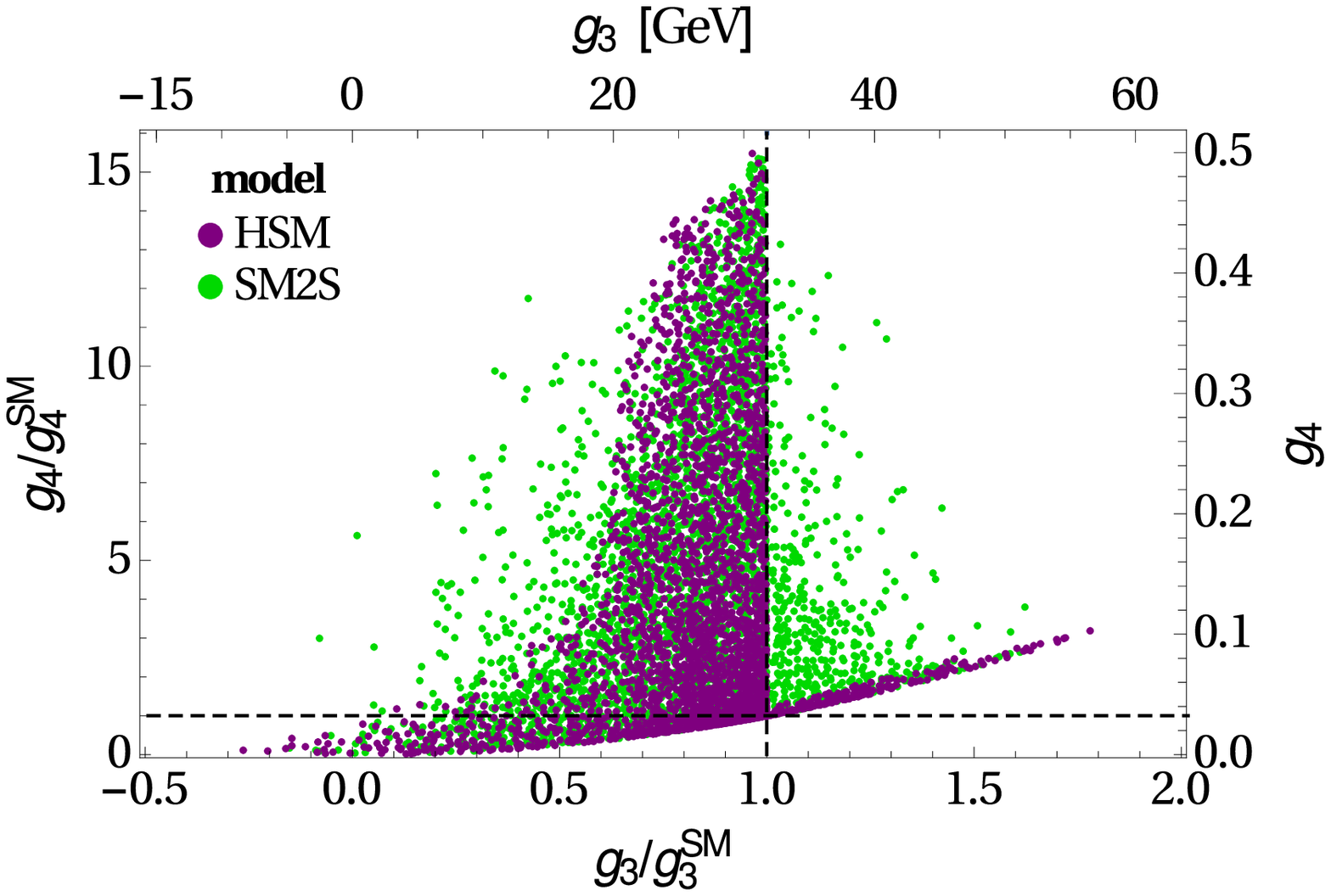,width=0.7\textwidth}
\end{center}
\caption{Scatter plot of $g_4 \left/ g_4^\mathrm{SM} \right.$ \textit{versus} 
  $g_3 \left/ g_3^\mathrm{SM} \right.$ in the HSM and in the SM2S.
  The dashed lines mark the Standard Model values
  $g_3 \left/ g_3^\mathrm{SM} \right. = g_4 \left/ g_4^\mathrm{SM} \right. = 1$.
  \label{fig42}}
\end{figure}

We assume that $\phi_1$ has VEV $v$ and $S$ has VEV $w$.
We write $S = w + \sigma$ together with equation~\eqref{nuiho}.
The mass matrix for $H$ and $\sigma$ is
\be
\left( \begin{array}{cc} 2 \lambda v^2 & 2 \sqrt{2} \xi v w \\
  2 \sqrt{2} \xi v w & 4 \psi w^2 \end{array} \right)
= \left( \begin{array}{cc} c & -s \\ s & c \end{array} \right)
\left( \begin{array}{cc} M_1 & 0 \\ 0 & M_2 \end{array} \right)
\left( \begin{array}{cc} c & s \\ -s & c \end{array} \right),
\ee
where $c \equiv \cos{\vartheta}$ and $s \equiv \sin{\vartheta}$.
We assume $|c| > 0.9$.
The oblique parameter
\be
T = \frac{3 s^2}{16 \pi s_w^2 m_W^2} \left[
F \left( M_1, m_W^2 \right) - F \left( M_1, m_Z^2 \right)
- F \left( M_2, m_W^2 \right) + F \left( M_2, m_Z^2 \right) \right]
\ee
must satisfy $-0.04 < T < 0.20$.
The three- and four-Higgs couplings are given by
\bs
\ba
\frac{g_3}{g_3^\mathrm{SM}} &=& c^3 + \frac{\sqrt{2} v}{w}\, s^3,
\\
g_4 &=& \frac{\lambda}{8}\, c^4 + \frac{\psi}{2}\, s^4
+ \frac{\xi}{2}\, c^2 s^2.
\ea
\es
In figure~\ref{fig42} we compare the predictions of the HSM and of the SM2S
for $g_3$ and $g_4$.
One sees that there is no substantial difference between the two models.

\setcounter{equation}{0}
\renewcommand{\theequation}{B\arabic{equation}}

\section{Other stability points of the SM2S potential}
\label{appendixB}

In this appendix we consider more carefully
the various stability points of the potential of the SM2S
in equation~\eqref{potpotpot}.
The vacuum value of that potential is given by
\bs
\label{1}
\ba
V_0 \equiv \left\langle 0 \left| V \right| 0 \right\rangle
&=& \mu_1 v^2 + m_1^2 w_1^2 + m_2^2 w_2^2 \\
& & + \frac{\lambda_1 v^4}{2} + \frac{\psi_1 w_1^4}{2}
+ \frac{\psi_2 w_2^4}{2} \\
& & + \psi_3 w_1^2 w_2^2
+ \xi_1 v^2 w_1^2
+ \xi_2 v^2 w_2^2.
\ea
\es
Equations~\eqref{20} follow from the assumption that $v$,
$w_1$,
and $w_2$ are not zero.
Defining
\be
\label{det}
d \equiv \lambda_1 \psi_1 \psi_2
+ 2 \psi_3 \xi_1 \xi_2
- \lambda_1 \psi_3^2
- \psi_1 \xi_2^2
- \psi_2 \xi_1^2,
\ee
one obtains
\bs
\label{6}
\ba
V_0 &=&
- \frac{\lambda_1 v^4}{2} - \frac{\psi_1 w_1^4}{2}
- \frac{\psi_2 w_2^4}{2} - \psi_3 w_1^2 w_2^2
- \xi_1 v^2 w_1^2
- \xi_2 v^2 w_2^2
\label{kdpdsp} \\
&=& \frac{1}{2 d} \left[
  \left( \psi_3^2 - \psi_1 \psi_2 \right) \mu_1^2
  + \left( \xi_2^2 - \lambda_1 \psi_2 \right) \left( m_1^2 \right)^2
  + \left( \xi_1^2 - \lambda_1 \psi_1 \right) \left( m_2^2 \right)^2
  \right. \hspace*{8mm} \\ & &
  + 2 \left( \xi_1 \psi_2 - \psi_3 \xi_2 \right) \mu_1 m_1^2
  + 2 \left( \xi_2 \psi_1 - \psi_3 \xi_1 \right) \mu_1 m_2^2
  \\ & & \left.
  + 2 \left( \lambda_1 \psi_3 - \xi_1 \xi_2 \right) m_1^2 m_2^2 \right].
\ea
\es

The mass matrix $M$ of the scalars is real and symmetric
and is given in equation~\eqref{emghty}.
We assume that $M$ has three positive eigenvalues $M_1$,
$M_2$,
and $M_3$.
It follows that
all the principal minors of $M$ are positive.\footnote{The principal minors
  of a square matrix are the determinants of its principal submatrices.}
(This is called `Sylvester's criterion'~\cite{sylvester}.)
Thus,
\bs
\label{7}
\ba
\lambda_1 &>& 0, \label{7a} \\
\psi_1 &>& 0, \label{7b} \\
\psi_2 &>& 0, \label{7c} \\
\lambda_1 \psi_1 - \xi_1^2 &>& 0, \label{7d} \\
\lambda_1 \psi_2 - \xi_2^2 &>& 0, \label{7e} \\
\psi_1 \psi_2 - \psi_3^2 &>& 0, \label{7f} \\
d &>& 0. \label{7g}
\ea
\es
These inequalities display some resemblance to the BFB conditions~\eqref{BFB1},
\eqref{BFB2}.

We now consider other stability points of the potential
where either $v$ or $w_1$ or $w_2$ vanish.
\begin{enumerate}
\item There is a stability point where $w_1 = w_2 = 0$.
  At that point the potential has the value
  \be
  V^{(1)} \equiv - \frac{\mu_1^2}{2 \lambda_1}.
  \ee
\item Similarly,
  there are stability points where either $v = w_1 = 0$ or $v = w_2 = 0$.
  At those two points the values of the potential are,
  respectively,
  \bs
  \ba
  V^{(2)} &\equiv& - \frac{\left( m_2^2 \right)^2}{2 \psi_2}, \\
  V^{(3)} &\equiv& - \frac{\left( m_1^2 \right)^2}{2 \psi_1}.
  \ea
  \es
\item There is a stability point of the potential where $v = 0$
  but $w_1$ and $w_2$ are nonzero.
  At that point the potential takes the value
  \be
  V^{(4)} \equiv \frac{- \psi_2 \left( m_1^2 \right)^2
    - \psi_1 \left( m_2^2 \right)^2 + 2 \psi_3 m_1^2 m_2^2}{2
    \left( \psi_1 \psi_2 - \psi_3^2 \right)}.
  \ee
\item Similarly,
  there is a stability point where $w_1 = 0$ but $v \neq 0$ and $w_2 \neq 0$.
  At that point the value of the potential is
  \be
  V^{(5)} \equiv \frac{- \psi_2 \mu_1^2
    - \lambda_1 \left( m_2^2 \right)^2 + 2 \xi_2 \mu_1 m_2^2}{2
    \left( \lambda_1 \psi_2 - \xi_2^2 \right)}.
  \ee
\item Finally,
  there is another stability point with value
  \be
  V^{(6)} \equiv \frac{- \psi_1 \mu_1^2
    - \lambda_1 \left( m_1^2 \right)^2 + 2 \xi_1 \mu_1 m_1^2}{2
    \left( \lambda_1 \psi_1 - \xi_1^2 \right)}
  \ee
  of the potential.
\end{enumerate}

From inequalities~\eqref{7c} and~\eqref{7f} it follows that
$V^{(4)} \le V^{(2)}$ is equivalent to 
\be
\psi_2 \left[ - \psi_2 \left( m_1^2 \right)^2
  - \psi_1 \left( m_2^2 \right)^2 + 2 \psi_3 m_1^2 m_2^2 \right]
\le \left( \psi_3^2 - \psi_1 \psi_2 \right) \left( m_2^2 \right)^2,
\ee
which in turn is equivalent to
\be
- \left( \psi_2 m_1^2 - \psi_3 m_2^2 \right)^2 \le 0,
\ee
and this is obvioulsy true.
One thus concludes that $V^{(4)}$ can never be larger than $V^{(2)}$.
In similar fashion one finds that
\bs
\ba
V^{(4)} &\le& V^{(2)}, \\
V^{(4)} &\le& V^{(3)}, \\
V^{(5)} &\le& V^{(1)}, \\
V^{(5)} &\le& V^{(2)}, \\
V^{(6)} &\le& V^{(1)}, \\
V^{(6)} &\le& V^{(3)}.
\ea
\es

Next consider the inequality $V_0 \le V^{(4)}$.
Because of~\eqref{7f} and~\eqref{7g},
it is equivalent to
\bs
\label{1616}
\ba
\left( \psi_1 \psi_2 - \psi_3^2 \right)
\left[ \left( \psi_3^2 - \psi_1 \psi_2 \right) \mu_1^2
  + \left( \xi_2^2 - \lambda_1 \psi_2 \right) \left( m_1^2 \right)^2
  \right. & &
  \\
  + \left( \xi_1^2 - \lambda_1 \psi_1 \right) \left( m_2^2 \right)^2
  + 2 \left( \xi_1 \psi_2 - \psi_3 \xi_2 \right) \mu_1 m_1^2
  & &
  \\
  \left.
  + 2 \left( \xi_2 \psi_1 - \psi_3 \xi_1 \right) \mu_1 m_2^2
  + 2 \left( \lambda_1 \psi_3 - \xi_1 \xi_2 \right) m_1^2 m_2^2 \right] &\le&
d \left[ - \psi_2 \left( m_1^2 \right)^2
  \right. \hspace*{7mm}
  \\
  & &
  - \psi_1 \left( m_2^2 \right)^2
  \\
  & & \left.
  + 2 \psi_3 m_1^2 m_2^2 \right].
\ea
\es
Introducing the expression for $d$ in equation~\eqref{det},
one finds that the inequality~\eqref{1616} is equivalent to
\bs
\label{17}
\ba
\left( \psi_1 \psi_1 - \psi_3^2 \right) \left[
  \left( \psi_3^2 - \psi_1 \psi_2 \right) \mu_1^2
  + 2 \left( \xi_1 \psi_2 - \psi_3 \xi_2 \right) \mu_1 m_1^2
  \right. & &
  \\
  \left.
  + 2 \left( \xi_2 \psi_1 - \psi_3 \xi_1 \right) \mu_1 m_2^2 \right]
- \left( m_1^2 \right)^2  \left( \psi_2 \xi_1 - \psi_3 \xi_2 \right)^2
& &
\\
- \left( m_2^2 \right)^2 \left( \psi_1 \xi_2 - \psi_3 \xi_1 \right)^2
- 2 m_1^2 m_2^2 \left( \psi_2 \xi_1 - \psi_3 \xi_2 \right)
\left( \psi_1 \xi_2 - \psi_3 \xi_1 \right) &\le& 0.
\ea
\es
This may be written as
\be
\left[ \left( \psi_3^2 - \psi_1 \psi_2 \right) \mu_1
  + \left( \psi_2 \xi_1 - \psi_3 \xi_2 \right) m_1^2
  + \left( \psi_1 \xi_2 - \psi_3 \xi_1 \right) m_2^2 \right]^2 \ge 0,
\ee
which is of course true.
In similar fashion one obtains that
\bs
\ba
V_0 &\le& V^{(4)}, \\
V_0 &\le& V^{(5)}, \\
V_0 &\le& V^{(6)}.
\ea
\es

We have thus demonstrated that,
because of our assumption that all three eigenvalues of the matrix $M$
are positive,
$V_0$ is smaller than $V^{1,2,3,4,5,6}$,
\textit{viz.}\ the stability point of $V$ with nonzero $v$,
$w_1$,
and $w_2$ is the vacuum.

This result may be easily understood in the following way.
The potential~\eqref{potpotpot} of the SM2S may be rewritten
\be
\label{dnkfps}
V = \frac{1}{2}\, X^T \Lambda X + V_0,
\ee
where $V_0$ is the vacuum expectation value of the potential
given in equation~\eqref{kdpdsp} and
\be
X = \left( \begin{array}{c}
  \phi_1^\dagger \phi_1 - v^2 \\ S_1^2 - w_1^2 \\ S_2^2 - w_2^2
\end{array} \right),
\quad
\Lambda = \left( \begin{array}{ccc}
  \lambda_1 & \xi_1 & \xi_2 \\ \xi_1 & \psi_1 & \psi_3 \\ \xi_2 & \psi_3 & \psi_2
\end{array} \right).
\ee
We assume that the point
$X = \left( \begin{array}{ccc} 0,\ 0,\ 0 \end{array} \right)^T$
is a local minimum of the potential $V$.
Then,
since the potential in equation~\eqref{dnkfps} is a quadratic form is $X$,
the point
$X = \left( \begin{array}{ccc} 0,\ 0,\ 0 \end{array} \right)^T$
must also be the \emph{global}\/ minimum of $V$.\footnote{We thank Igor Ivanov for presenting this argument to us.}

\setcounter{equation}{0}
\renewcommand{\theequation}{C\arabic{equation}}

\section{Global minimum conditions for the 2HDM1S}
\label{appendixC}

In the 2HDM1S,
we define $q_1 \equiv \phi_1^\dagger \phi_1$,
$q_2 \equiv \phi_2^\dagger \phi_2$,
$z \equiv \phi_1^\dagger \phi_2$,
$z^\ast \equiv \phi_2^\dagger \phi_1$,\footnote{Since we only analyze
  the potential at the classical level,
  we simplify the notation by treating the fields as $c$-numbers
  instead of $q$-numbers.}
and $q_3 \equiv S^2$.
Note that
\be
q_1 \ge 0, \quad q_2 \ge 0, \quad
\left| z \right|^2 \le q_1 q_2, \quad q_3 \ge 0.
\label{mgkffo}
\ee
We define the column vector $X = \left( \begin{array}{ccccc}
  q_1, & q_2, & z, & z^\ast, & q_3 \end{array} \right)^T$.
The scalar potential of the 2HDM1S may then be written as
\be
\label{bkfpdsos}
V = Y^T X + \frac{1}{2}\, X^T \Lambda\, X,
\ee
where
\be
Y = \left( \begin{array}{c}
  \mu_1 \\ \mu_2 \\ \mu_3 \\ \mu_3^\ast \\ \mu_4
\end{array} \right),
\quad
\Lambda = \left( \begin{array}{ccccc}
  \lambda_1 & \lambda_3 & \lambda_6 & \lambda_6^\ast & \xi_1 \\
  \lambda_3 & \lambda_2 & \lambda_7 & \lambda_7^\ast & \xi_2 \\
  \lambda_6 & \lambda_7 & \lambda_5 & \lambda_4 & \xi_3 \\
  \lambda_6^\ast & \lambda_7^\ast & \lambda_4 & \lambda_5^\ast & \xi_3^\ast \\
  \xi_1 & \xi_2 & \xi_3 & \xi_3^\ast & \psi
\end{array} \right).
\ee
The coefficients $\mu_1$,
$\mu_2$,
$\mu_3$,
and $\mu_4$ contained in the column vector $Y$ have squared-mass dimension;
$\mu_3$ is in general complex while $\mu_1$,
$\mu_2$,
and $\mu_4$ are real.
The coefficients contained in the symmetric matrix $\Lambda$
are treated by us as an input,
\textit{cf.}\ section~\ref{proc2HDM1S}.
Since we study the 2HDM1S in the Higgs basis,
where $\phi_2$ has zero VEV,
in the vacuum one has $q_2 = z = z^\ast = 0$,
$q_1 = v^2$,
and $q_3 = w^2$;
the vacuum expectation value of the potential is
\be
\label{gjgosps}
V_0 \equiv \left\langle 0 \left| V \right| 0 \right\rangle
= \mu_1 v^2 + \mu_4 w^2 + \frac{\lambda_1 v^4}{2} + \frac{\psi w^4}{2}
+ \xi_1 v^2 w^2.
\ee
It follows that
\bs
\label{mu1}
\ba
\mu_1 &=& - \lambda_1 v^2 - \xi_1 w^2, \\
\mu_4 &=& - \xi_1 v^2 - \psi w^2.
\ea
\es
Solving for $v^2$ and $w^2$ the system~\eqref{mu1}
and plugging the solution into equation~\eqref{gjgosps},
one obtains
\be
V_0 = \frac{- \psi \left( \mu_1 \right)^2 - \lambda_1 \left( \mu_4 \right)^2
  + 2 \xi_1 \mu_1 \mu_4}{2
  \left[ \psi \lambda_1 - \left( \xi_1 \right)^2 \right]}.
\label{nbjigho}
\ee
Moreover,
in the Higgs basis
\bs
\label{mu4}
\ba
\mu_2 &=& M_C - \lambda_3 v^2 - \xi_2 w^2, \label{vjifof} \\
\mu_3 &=& - \lambda_6 v^2 - \xi_3 w^2.
\ea
\es
In equation~\eqref{vjifof},
$M_C$ is the mass of the charged scalar;
we treat it as an input,
just as $v$ and $w$.\footnote{More exactly,
  we input $v = 174$\,GeV and the squared mass
  $M_1 = \left( 125\, \mathrm{GeV} \right)^2$ of one of the scalars,
  and we derive the value of $w$ therefrom.}
By using equations~\eqref{mu1} and~\eqref{mu4} we find the values of $\mu_1$,
$\mu_2$,
$\mu_3$,
and $\mu_4$ from the input.

We want to check that,
for each set of input parameters
(\textit{i.e.}\ $\lambda_{1,\ldots,7}$,
$\xi_{1,2,3}$,
$\psi$,
$v$,
$w$,
and $M_C$)
in our data set,
the state that we \emph{assume}\/ to be the vacuum,
characterized by $q_2 = z = z^\ast = 0$,
is \emph{indeed}\/ the \emph{global}\/ minimum of the potential.
In order to do this we must consider all the other possible stability points
of the potential
and check that the value of the potential at each of those points
is larger than $V_0$ in equation~\eqref{nbjigho}.
The stability points may either be inside the domain
defined by equations~\eqref{mgkffo}
or they may lie on a boundary of that domain.
There is only one possible stability point inside the domain;
deriving equation~\eqref{bkfpdsos} relative to $X$,
we find that it is given by
\bs
\ba
X \equiv X^{(1)} &=& - \Lambda^{-1} Y, \label{buihouy} \\
V \equiv V^{(1)} &=& - \frac{1}{2}\, Y^T \Lambda^{-1} Y. \label{hjugigt}
\ea
\es
For each set of input parameters,
we have computed the column vector $X^{(1)}$ by using equation~\eqref{buihouy}.
If that vector happened to be inside the domain,
\textit{viz.}\ if $X^{(1)}_1 > 0$,
$X^{(1)}_2 > 0$,
$\left| X^{(1)}_3 \right|^2 < X^{(1)}_1 X^{(1)}_2$,
and $X^{(1)}_4 > 0$,
then we computed $V^{(1)}$ by using equation~\eqref{hjugigt}.
We checked whether $V^{(1)} > V_0$;
if the latter condition did \emph{not}\/ hold,
then we discarded that set of input parameters.

Next we have considered the various possible stability points
on boundaries of the domain.
Firstly there is the boundary with $q_3 = 0$ but $q_1 > 0$,
$q_2 > 0$,
and $\left| z \right|^2 < q_1 q_2$.
In that case one has
\be
V = \bar Y^T \bar X + \frac{1}{2}\, \bar X^T \bar \Lambda\, \bar X,
\ee
where
\be
\bar X =
\left( \begin{array}{c} q_1 \\ q_2 \\ z \\ z^\ast \end{array} \right),
\quad
\bar Y =
\left( \begin{array}{c} \mu_1 \\ \mu_2 \\ \mu_3 \\ \mu_3^\ast
\end{array} \right),
\quad
\bar \Lambda = \left( \begin{array}{cccc}
  \lambda_1 & \lambda_3 & \lambda_6 & \lambda_6^\ast \\
  \lambda_3 & \lambda_2 & \lambda_7 & \lambda_7^\ast \\
  \lambda_6 & \lambda_7 & \lambda_5 & \lambda_4 \\
  \lambda_6^\ast & \lambda_7^\ast & \lambda_4 & \lambda_5^\ast
\end{array} \right).
\ee
There is one possible stability point with
\bs
\ba
\bar X \equiv \bar X^{(2)} &=& - \bar \Lambda^{-1} \bar Y,
\label{vjghdo}
\\
V \equiv V^{(2)} &=& - \frac{1}{2}\, \bar Y^T \bar \Lambda^{-1} \bar Y.
\label{vjghdos}
\ea
\es
For each set of input parameters,
we have computed the column vector $\bar X^{(2)}$
by using equation~\eqref{vjghdo}.
Whenever that vector happened to fulfil $\bar X^{(2)}_1 > 0$,
$\bar X^{(2)}_2 > 0$,
and $\left| \bar X^{(2)}_3 \right|^2 < \bar X^{(2)}_1 \bar X^{(2)}_2$,
we computed $V^{(2)}$ by using equation~\eqref{vjghdos}.
We checked whether $V^{(2)} > V_0$;
if that condition did not hold,
then we discarded the set of input parameters.

Secondly we have checked a possible stability point
with null $q_1$ (and $z$) instead of null $q_2$ (and $z$).
In analogy with equations~\eqref{mu1} and~\eqref{nbjigho},
in that case one has
\bs
\ba
q_2 &=& \frac{- \psi \mu_2 + \xi_2 \mu_4}{\psi \lambda_2
  - \left( \xi_2 \right)^2},
\label{qq2} \\
q_3 &=& \frac{\xi_2 \mu_2 - \lambda_2 \mu_4}{\psi \lambda_2
  - \left( \xi_2 \right)^2},
\label{qq4} \\
V \equiv V^{(3)}
&=& \frac{- \psi \left( \mu_2 \right)^2 - \lambda_2 \left( \mu_4 \right)^2
  + 2 \xi_2 \mu_2 \mu_4}{2
  \left[ \psi \lambda_2 - \left( \xi_2 \right)^2 \right]}.
\label{vvs}
\ea
\es
For each set of parameters,
we have computed $q_2$ and $q_3$ through equations~\eqref{qq2} and~\eqref{qq4},
respectively.
Whenever $q_2$ and $q_3$ were both positive,
we have computed $V^{(3)}$ through equation~\eqref{vvs};
if $V^{(3)} < V_0$,
then we discarded the set of parameters.

Thirdly,
we have considered the following possible stability points on boundaries
of the domain:
\begin{enumerate}
\item The point $q_1 = q_2 = z = q_3 = 0$ has $V = 0$,
  Therefore,
  when $V_0 > 0$ we have discarded the set of parameters.
\item When $q_1 = q_2 = z = 0$ but $q_3 \neq 0$,
  there is a stability point featuring
  \bs
  \ba
  q_3 &=& - \frac{\mu_4}{\psi}, \label{hjigov} \\
  V \equiv V^{(4)} &=& - \frac{\left( \mu_4 \right)^2}{2 \psi}. \label{bjigdc}
  \ea
  \es
  Whenever $q_3$ in equation~\eqref{hjigov} happened to be positive
  and simultaneously $V^{(4)}$ in equation~\eqref{bjigdc} was smaller then $V_0$,
  we have discarded the set of parameters.
\item When $q_1 = q_3 = z = 0$ but $q_2 \neq 0$,
  there is a stability point featuring
  \bs
  \ba
  q_2 &=& - \frac{\mu_2}{\lambda_2}, \label{hjigov2} \\
  V \equiv V^{(5)} &=&
  - \frac{\left( \mu_2 \right)^2}{2 \lambda_2}. \label{bjigdc2}
  \ea
  \es
  Whenever $q_2$ in equation~\eqref{hjigov2} happened to be positive
  and simultaneously $V^{(5)}$ in equation~\eqref{bjigdc2}
  was smaller then $V_0$,
  we have discarded the set of parameters.
\item When $q_2 = q_3 = z = 0$ but $q_1 \neq 0$,
  there is a stability point featuring
  \bs
  \ba
  q_1 &=& - \frac{\mu_1}{\lambda_1}, \label{hjigov3} \\
  V \equiv V^{(6)} &=&
  - \frac{\left( \mu_1 \right)^2}{2 \lambda_1}. \label{bjigdc3}
  \ea
  \es
  Whenever $q_1$ in equation~\eqref{hjigov3} happened to be positive
  and simultaneously $V^{(6)}$ in equation~\eqref{bjigdc3}
  was smaller then $V_0$,
  we have discarded the set of parameters.
\end{enumerate}

All the above tests are easily applied.
The awkward tests involve the boundaries where $\left| z \right|^2 = q_1 q_2$.
In that case one writes $z = e^{i \theta} \sqrt{q_1 q_2}$ to obtain
\bs
\label{hcuiyop}
\ba
V \equiv \hat V_0 &=&
\mu_1 q_1 + \mu_2 q_2
+ 2\, \Re{\left( \mu_3 e^{i \theta} \right)} \sqrt{q_1 q_2}
+ \mu_4 q_3
\\ & &
+ \frac{\lambda_1}{2} \left( q_1 \right)^2
+ \frac{\lambda_2}{2} \left( q_2 \right)^2
+ \frac{\psi}{2} \left( q_3 \right)^2
+ \left[ \lambda_3 + \lambda_4
  + \Re{\left( \lambda_5 e^{2 i \theta} \right)} \right] q_1 q_2
\\ & &
+ 2\, \Re{\left( \lambda_6 e^{i \theta} \right)} q_1 \sqrt{q_1 q_2}
+ 2\, \Re{\left( \lambda_7 e^{i \theta} \right)} q_2 \sqrt{q_1 q_2}
\\ & &
+ \xi_1 q_1 q_3
+ \xi_2 q_2 q_3
+ 2\, \Re{\left( \xi_3 e^{i \theta} \right)} q_3 \sqrt{q_1 q_2}.
\ea
\es
Deriving $\hat V_0$ in equation~\eqref{hcuiyop} relative to $q_1$,
$q_2$,
$q_3$,
and $\theta$ one obtains the stability equations
\bs
\label{system2}
\ba
0 &=& \mu_1
+ \Re{\left( \mu_3 e^{i \theta} \right)} \sqrt{\frac{q_2}{q_1}}
+ \lambda_1 q_1
+ \left[ \lambda_3 + \lambda_4
  + \Re{\left( \lambda_5 e^{2 i \theta} \right)} \right] q_2
\\ & &
+ 3\, \Re{\left( \lambda_6 e^{i \theta} \right)} \sqrt{q_1 q_2}
+ \Re{\left( \lambda_7 e^{i \theta} \right)} q_2 \sqrt{\frac{q_2}{q_1}}
\\ & &
+ \xi_1 q_3
+ \Re{\left( \xi_3 e^{i \theta} \right)} q_3 \sqrt{\frac{q_2}{q_1}},
\\*[1mm]
0 &=& \mu_2
+ \Re{\left( \mu_3 e^{i \theta} \right)} \sqrt{\frac{q_1}{q_2}}
+ \lambda_2 q_2
+ \left[ \lambda_3 + \lambda_4
  + \Re{\left( \lambda_5 e^{2 i \theta} \right)} \right] q_1
\\ & &
+ \Re{\left( \lambda_6 e^{i \theta} \right)} q_1 \sqrt{\frac{q_1}{q_2}}
+ 3\, \Re{\left( \lambda_7 e^{i \theta} \right)} \sqrt{q_1 q_2}
\\ & &
+ \xi_2 q_3
+ \Re{\left( \xi_3 e^{i \theta} \right)} q_3 \sqrt{\frac{q_1}{q_2}},
\\*[1mm]
0 &=& \mu_4 + \psi q_3 + \xi_1 q_1 + \xi_2 q_2
+ 2\, \Re{\left( \xi_3 e^{i \theta} \right)} \sqrt{q_1 q_2},
\\*[1mm]
0 &=&
\Im{\left( \mu_3 e^{i \theta} \right)}
+ \Im{\left( \lambda_5 e^{2 i \theta} \right)} \sqrt{q_1 q_2}
\\ & &
+ \Im{\left( \lambda_6 e^{i \theta} \right)} q_1
+ \Im{\left( \lambda_7 e^{i \theta} \right)} q_2
+ \Im{\left( \xi_3 e^{i \theta} \right)} q_3.
\ea
\es
For each set of parameters of the potential,
we have searched for solutions,
\textit{i.e.}\ for $q_1 > 0$,
$q_2 > 0$,
$q_3 > 0$,
and a phase $\theta$
satisfying the system~\eqref{system2} of four equations.
(This proved to be a highly nontrivial task.)
Whenever we found a solution,
we computed $\hat V_0$ through equation~\eqref{hcuiyop}
and checked whether $\hat V_0 < V_0$;
when that happened for at least one solution of~\eqref{system2},
we have discarded the corresponding set of parameters.

One must also consider the domain border
$\left| z \right|^2 = q_1 q_2$ and $q_3 = 0$.
In that case one must solve the simpler system of equations
\bs
\label{system}
\ba
0 &=& \mu_1
+ \Re{\left( \mu_3 e^{i \theta} \right)} \sqrt{\frac{q_2}{q_1}}
+ \lambda_1 q_1
+ \left[ \lambda_3 + \lambda_4
  + \Re{\left( \lambda_5 e^{2 i \theta} \right)} \right] q_2
\\ & &
+ 3\, \Re{\left( \lambda_6 e^{i \theta} \right)} \sqrt{q_1 q_2}
+ \Re{\left( \lambda_7 e^{i \theta} \right)} q_2 \sqrt{\frac{q_2}{q_1}},
\\*[1mm]
0 &=& \mu_2
+ \Re{\left( \mu_3 e^{i \theta} \right)} \sqrt{\frac{q_1}{q_2}}
+ \lambda_2 q_2
+ \left[ \lambda_3 + \lambda_4
  + \Re{\left( \lambda_5 e^{2 i \theta} \right)} \right] q_1
\\ & &
+ \Re{\left( \lambda_6 e^{i \theta} \right)} q_1 \sqrt{\frac{q_1}{q_2}}
+ 3\, \Re{\left( \lambda_7 e^{i \theta} \right)} \sqrt{q_1 q_2},
\\*[1mm]
0 &=&
\Im{\left( \mu_3 e^{i \theta} \right)}
+ \Im{\left( \lambda_5 e^{2 i \theta} \right)} \sqrt{q_1 q_2}
+ \Im{\left( \lambda_6 e^{i \theta} \right)} q_1
+ \Im{\left( \lambda_7 e^{i \theta} \right)} q_2.
\ea
\es
For each set of parameters,
whenever we found a solution $q_1 > 0$,
$q_2 > 0$,
and $\theta$ of equations~\eqref{system} we computed
\bs
\ba
\tilde V_0 &=&
\mu_1 q_1 + \mu_2 q_2
+ 2\, \Re{\left( \mu_3 e^{i \theta} \right)} \sqrt{q_1 q_2}
\\ & &
+ \frac{\lambda_1}{2} \left( q_1 \right)^2
+ \frac{\lambda_2}{2} \left( q_2 \right)^2
+ \left[ \lambda_3 + \lambda_4
  + \Re{\left( \lambda_5 e^{2 i \theta} \right)} \right] q_1 q_2
\\ & &
+ 2\, \Re{\left( \lambda_6 e^{i \theta} \right)} q_1 \sqrt{q_1 q_2}
+ 2\, \Re{\left( \lambda_7 e^{i \theta} \right)} q_2 \sqrt{q_1 q_2}.
\ea
\es
If $\tilde V_0 < V_0$ for any solution of equations~\eqref{system},
then we discarded the set of input parameters.

By applying all the tests in this appendix,
we have eliminated about half of our initial set of sets of input parameters.
Thus,
the tests in this appendix prove crucial in the correct analysis
of the 2HDM1S.

We have also applied the tests in this appendix,
with the necessary simplifications,
to the case of the 2HDM~\cite{Xu:2017vpq}.
In particular,
in that case we do not have to solve
the very complicated system of four equations~\eqref{system2},
we only have to solve the much easier system of three equations~\eqref{system}.
We have checked that the tests in this appendix yield,
for the 2HDM, exactly the same result as the much simpler method
described in the paragraph between equations~\eqref{T_2HDM} and~\eqref{g4_2HDM}.

\end{appendix}

\end{document}